\documentclass[useAMS,usenatbib]{mn2e}
\usepackage{epsfig}
\usepackage{subfigure}
\usepackage{graphicx}

\def\gsim{\;\lower4pt\hbox{${\buildrel\displaystyle >\over\sim}$}\;}
\def\lsim{\;\lower4pt\hbox{${\buildrel\displaystyle <\over\sim}$}\;}
\def\grls{\;\lower4pt\hbox{${\buildrel\displaystyle >\over <}$}\;}
\title[Similarity Polytropic Gas Dynamics]
{New Self-Similar Solutions of Polytropic Gas Dynamics}
\author[Y.-Q. Lou \& W.-G. Wang]{Yu-Qing Lou$^{1,2,3}$
\thanks{E-mail: louyq@tsinghua.edu.cn and lou@oddjob.uchicago.edu;
wwg03@mails.tsinghua.edu.cn } and Wei-Gang Wang$^{1}$
\\
$^{1}$Physics Department and Tsinghua Center for Astrophysics
(THCA), Tsinghua University, Beijing, 100084, China;\\
$^{2}$Department of Astronomy and Astrophysics, the University
of Chicago, 5640 South Ellis Avenue, Chicago, IL 60637, USA;\\
$^{3}$National Astronomical Observatories, Chinese Academy of
Sciences, A20, Datun Road, Beijing 100012, China. \\}

\begin{document}

\date{Accepted 2006 ?? ??. Received 2006 ?? ??;
in original form 2006 March ??}

\pagerange{\pageref{firstpage}--\pageref{lastpage}}\pubyear{2006}

\maketitle

\label{firstpage}

\begin{abstract}
We explore semi-complete self-similar solutions for the polytropic
gas dynamics involving self-gravity under spherical symmetry,
examine behaviours of the sonic critical curve, and present
new asymptotic collapse solutions that describe `quasi-static'
asymptotic behaviours at small radii and large times. These new
`quasi-static' solutions with divergent mass density approaching
the core can have self-similar oscillations.
%in both the first-order velocity field and the second-order density field.
Earlier known solutions are summarized.
Various semi-complete self-similar solutions involving such novel
asymptotic solutions are constructed, either with or without a shock.
In contexts of stellar core collapse and supernova explosion, a
hydrodynamic model of a rebound shock initiated around the stellar
degenerate core of a massive progenitor star is presented. With
this dynamic model framework, we attempt to relate progenitor
stars and the corresponding remnant compact stars: neutron stars,
black holes, and white dwarfs.
\end{abstract}

\begin{keywords}
black hole physics --- hydrodynamics ---
%planetary nebulae: general--- stars: AGB and post-AGB
%--- stars: formation --- stars: winds, outflows ---
shock waves --- stars: neutron --- supernovae --- white dwarfs
\end{keywords}

\section[]{Introduction}

The self-similar evolution of gas dynamics with spherical symmetry
under self-gravity and thermal pressure has attracted considerable
interest in contexts of various astrophysical applications
[\cite{shu1977}; \cite{goldreich80}; \cite{fillmore84}; \cite{lou2005};
\cite{HSLZ2006}].
%and extensive references therein].
There are several reasons to pursue such model studies: numerical
simulations show the tendency of self-similar behaviours in spherical
flows (e.g. Bodenheimer \& Sweigart 1968 and Foster \& Chevalier 1993);
this model framework is sufficiently simple, yet can still carry
valuable information on the roles of the competing forces of gravity
and thermal pressure. Larson (1969a, b) and Penston (1969a, b) were
the first to analyse such similarity problems in astrophysical contexts.
Since then, different asymptotic solutions and global numerical
solutions have been found and developed. In the isothermal case,
\cite{shu1977} obtained the inner free-fall asymptotic solution
with divergent speed and density profiles at small $x$, and
constructed the expansion wave collapse solution (EWCS) to suggest
the inside-out collapse scenario for the formation of low-mass
stars. Subsequently, \cite{hunter1977} discovered a sequence of
discrete solutions, referred to as the Hunter type solutions,
which have finite density and speed profiles at small radii,
by a matching procedure in the density-speed phase diagram.
\cite{whitworthsummers1985} noted the existence of asymptotic
solutions of constant speed at large radii, as a generalization
of the asymptotic solution at large $x$ of Shu (1977), and
constructed solutions with weak discontinuities across the sonic
critical line. The properties of such weak discontinuities were
further discussed by \cite{hunter1986}. Recently, \cite{loushen2004}
constructed isothermal solutions to describe self-similar evolution
of envelope expansion with core collapse (EECC) using the similar
matching procedure of \cite{hunter1977}. Bian \& Lou (2005)
explored various similarity isothermal shock solutions.

For a polytropic gas, \cite{cheng78} introduced a generalized
self-similar transformation in the sense of $\kappa\equiv p/
\rho^{\gamma}$ being constant along each streamline but not
globally, where $1\leq\gamma\leq5/3$ and the initial density
profile may be adjusted, and mainly studied the polytropic
counterparts of isothermal EWCSs. \cite{yahil83} developed a
model for $\kappa$ being a global constant, discussed the
polytropic case for $6/5\leq\gamma\leq4/3$, and mainly focussed
on the polytropic counterparts of Hunter-type isothermal solutions.
\cite{sutosilk88} used a similarity transformation similar to that
of \cite{yahil83}, but discussed a generalization of the equation
of state for $1<\gamma<4/3$ involving yet another parameter $n$,
and mainly constructed the counterparts of isothermal free-fall
solutions both crossing and not crossing the sonic critical curve;
\cite{mclaughlin97} treated the limit case of $\gamma\rightarrow 0$
using the logotropic equation of state, and considered the free-fall
solutions and EWCSs as counterparts of earlier analyses.
\cite{fatuzzo2004} explored the possibility of an initial equation
of state being different from a later dynamic equation of state, and
examined the dependence of the accreted mass versus the polytropic
index; their self-similar solutions do not encounter the sonic critical
curve. The basic polytropic model framework has also been extended in
various relevant physical aspects, e.g., \cite{tsc} for a slowly
rotating gas cloud, \cite{boily95} for including various forms of
radiative losses, \cite{semelin01} for including viscosity,
\cite{WL06} for modelling a random magnetic field, as well as other
relevant research works involving hydrodynamic and MHD shocks
\citep{BianLou05, YuLou06}.

Among these earlier research works, \cite{sutosilk88} introduced a
straightforward self-similar transformation and analyzed polytropic
self-similar flows as a generalization of an isothermal gas (e.g.
Shu 1977). They analyzed both cases of $\gamma>1$, $n=1$ and
of $n=2-\gamma$. The $n=1$ case is somewhat artificial because this
equation of state evolves with time and does not conserve the local
specific entropy. We therefore focus on
%take the formulation and notations of \cite{sutosilk88} but only for
the $n=2-\gamma$ case in this paper to explore a polytropic model for
the collapse of a massive star. As shocks are ubiquitous and important
in various astrophysical processes (e.g. Kennel \& Coroniti 1984 for
an MHD model of the Crab Nebula with shocks), and self-similar shocks
have been investigated in various astrophysical contexts (e.g. Tsai
\& Hsu 1995; Shu et al. 2002; Shen \& Lou 2004; Bian \& Lou 2005; Yu,
Lou, Bian \& Wu 2006), we further construct self-similar shocks in a
polytropic gas in this paper. We analyze the behaviour of the sonic
critical curve and present new asymptotic solution referred to as the
`quasi-static' asymptotic solution. Semi-complete solutions containing
the `quasi-static' asymptotic solution as the limit are constructed,
either with or without a shock. We invoke this simple theoretical
framework to model the collapse of a massive progenitor star involving
a rebound shock in the stellar interior, expelling stellar materials
and leaving behind a remnant compact object, such as white dwarfs,
neutron stars, and black holes.

This paper is arranged as follows. The general background
information in provided in Section 1. Section 2 gives an account
for the hydrodynamic formulation of the model problem
%in parallel with that of \cite{sutosilk88}, as well as
and previously known solutions. Section 3 treats the sonic critical
curve. In Section 4, we present the novel `quasi-static' asymptotic
solutions. We describe self-similar shocks and the relevant jump
condition in Section 5. In Section 6, we show various semi-complete
solutions either with or without a shock. Section 7 contains
numerical examples for a rebound shock within a collapsing star.
Finally in Section 8, we provide summary and conclusions.

\section[]{Formulation and Known Solutions}

\subsection[]{Nonlinear Hydrodynamic Equations}

The present theoretical model problem treats the self-gravitational
fluid dynamics under the spherical symmetry. Our analysis and
results are applicable to the vast region sufficiently far away
from the central sphere surrounding various core activities. In
the spherical polar coordinates $(r, \theta, \phi)$, the standard
nonlinear polytropic hydrodynamic equations are in the forms of
\begin{equation}\label{mass1}
\frac{\partial \rho}{\partial t}
+\frac{1}{r^{2}}\frac{\partial}{\partial r}(r^{2}\rho u)=0\ ,
\end{equation}

\begin{equation}\label{mass2}
\frac{\partial M}{\partial t}+u\frac{\partial M}{\partial r}=0\ ,
\end{equation}

\begin{equation}\label{mass3}
\frac{\partial M}{\partial r}=4\pi r^{2}\rho\ ,
\end{equation}

\begin{equation}\label{momentum}
\rho\bigg(\frac{\partial u}{\partial t}
+u\frac{\partial u}{\partial r}\bigg)
=-\frac{\partial p}{\partial r}
-\frac{GM\rho}{r^{2}}\ ,
\end{equation}

\begin{equation}\label{state}
p=\kappa\rho^{\gamma}\ ,
\end{equation}
where $u(r,t)$ is the radial bulk fluid velocity at radius $r$ and
time $t$, $\rho(r,t)$ is the gas  mass density, $M(r,t)$ is the total
mass enclosed within radius $r$ at time $t$, $G=6.67\times 10^{-8}
\hbox{ g}^{-1}\hbox{ cm}^{-3}\hbox{ s}^{-2}$ is the gravitational
constant, and $p$ is the thermal gas pressure. The Poisson equation
relating the gravitational potential $\Phi$ (such that $d\Phi/dr=
GM/r^2$) and $\rho$ is consistently satisfied. In equation of state
(\ref{state}), the proportional coefficient $\kappa$ is independent
of $t$ and is assumed to be a global constant. This $\kappa$ would
be the same as the notation $K(t)$ in equation (11) of
\cite{sutosilk88} when the parameter $n$ of \cite{sutosilk88} (and
also in this paper to be defined presently) is equal to $2-\gamma$.
The range $1<\gamma<4/3$ for the polytropic index $\gamma$ is
adopted in our current model analysis.\footnote{See Appendix \ref{gamma}
for a brief discussion on cases of $\gamma\geq4/3$.}

\subsection[]{A Self-Similar Transformation}

%In parallel with \cite{sutosilk88},
We introduce the self-similar transformation
\citep{sutosilk88} with respect to a new
dimensionless independent variable $x$ as follows.
\begin{equation}\label{transform1}
r\equiv ax\ ,\ u\equiv bv\ ,\ \rho\equiv c\alpha\ ,\
p\equiv d\beta\ ,\ M\equiv em\ ,
\end{equation}
where the scaling factors $a(t)$
through $e(t)$ are defined by
\[
\quad a\equiv k^{1/2}t^{n}\ ,\qquad b\equiv k^{1/2}t^{n-1}\ ,
\qquad c\equiv \frac{1}{4\pi Gt^{2}}\ ,
\]
\begin{equation}\label{transform2}
\quad d\equiv\frac{kt^{2n-4}}{4\pi G}\ ,\qquad
e\equiv\frac{k^{3/2}t^{3n-2}}{(3n-2)G}\ ,
\end{equation}
with $k$ and $n$ being two constants. Here $v(x)$,
$\alpha(x)$, $\beta(x)$ and $m(x)$ are dimensionless functions
of $x$ only. Substituting expressions (\ref{transform1}) and
(\ref{transform2}) into the original polytropic hydrodynamic
equations (\ref{mass1}) through (\ref{state}) and assuming
$n=2-\gamma$ for a global constant $\kappa=k(4\pi G)^{\gamma-1}$,
we obtain
%in a straightforward manner
\begin{equation}\label{ode1}
m=\alpha x^2(nx-v)\ ,
\end{equation}
\[
\alpha'=\alpha^{2}\bigg[(n-1)v+\frac {nx-v}{3n-2}\alpha
\]
\begin{equation}\label{ode2}
\qquad\qquad
-\frac {2(x-v)(nx-v)}{x}\bigg]\big[\alpha (nx-v)^{2}
-\gamma\alpha^{\gamma}\big]^{-1}\ ,
\end{equation}
and
\[
v'=\bigg[(n-1)\alpha v(nx-v)
+\frac{(nx-v)^{2}}{(3n-2)}\alpha^2
\]
\begin{equation}\label{ode3}
\qquad\qquad
-2\gamma\alpha^{\gamma}\frac{(x-v)}{x}\bigg]
\big[\alpha (nx-v)^{2}-\gamma\alpha^{\gamma}\big]^{-1}\ .
\end{equation}
So far, the development is the same as that of \cite{sutosilk88}
with the requirement of $n=2-\gamma$. From equation (\ref{ode1}),
it is clear that $nx-v>0$ is necessary for $m>0$ as a physical
requirement for solutions.

\subsection[]{Known Similarity Solutions}

Various solutions of the two coupled nonlinear ordinary
differential equations (ODEs) (\ref{ode2}) and (\ref{ode3})
have been known previously. These solutions with $n=2-\gamma$,
which were first established in the isothermal case and then
generalized to the polytropic case, are summarized below.

The static solution of a singular polytropic
sphere (SPS) is characterized by
\[
v=0,\qquad
\alpha=\bigg[\frac{n^{2}}
{2\gamma(3n-2)}\bigg]^{1/n}x^{-2/n}\ ,
\]
\begin{equation}\label{pre1}
m=n\bigg[\frac{n^{2}} {2\gamma (3n-2)}\bigg]^{1/n}
x^{3-2/n}\ .
\end{equation}
This is to be compared with the singular isothermal
sphere (SIS; e.g., Shu 1977) and with the magnetized
singular isothermal sphere (mSIS; Yu \& Lou 2005).

The Larson-Penston type of solutions is generalized to
\begin{equation}\label{pre2}
v=\frac{2x}{3}\ ,\qquad  \alpha=\frac{2}{3}\ ,
\qquad m=\frac{2(n-2/3)}{3}x^{3}\ ,
\end{equation}
which was found by Larson (1969a, b) and Penston (1969a, b)
in the isothermal case with $n=1$; this is also known as the
non-relativistic Einstein de-Sitter polytropic expansion
solution (see Shu et al. 2002 for the isothermal counterpart).

The asymptotic solution finite at large $x$ is given by
\[
v=\bigg(-\frac{nA}{3n-2}+\frac{2\gamma
A^{\gamma-1}}{n}\bigg)x^{(n-2)/n}+Bx^{(n-1)/n}\ ,
\]
\begin{equation}\label{pre3}
\alpha=Ax^{-2/n}\ ,
\end{equation}
where $A$ and $B$ are two constants of integration. The
isothermal counterpart solution with $B=0$ and $n=1$
was discussed by \cite{shu1977}, and the more general
solution containing the free parameter $B$ was first
obtained by \cite{whitworthsummers1985}. In this paper,
asymptotic solutions with $B=0$ and $v>0$ as the leading
term at large $x$ is referred to as breeze solutions,
asymptotic solutions with $B=0$ and $v<0$ as the leading
term at large $x$ is referred to as the contraction
solution, asymptotic solutions with $B>0$ are referred
to as outflow or wind solutions and asymptotic solutions
with $B<0$ are referrred to as inflow solutions.

The leading diverging behaviour of central free-fall
collapse solution in a polytropic gas at small $x$ is
characterized by
\[
v=-\bigg[\frac{2m(0)}{(3n-2)x}\bigg]^{1/2}\ ,
\]
\begin{equation}\label{pre4}
\alpha=\bigg[\frac{(3n-2)m(0)}{2x^{3}}\bigg]^{1/2}\ ,
\end{equation}
with $m(x)$ approaching a finite value $m(0)$, which
was first found by \cite{shu1977} for the isothermal
case with $n=1$.

The Hunter type of asymptotic solutions
at small $x$ are given by
\[
v=\frac{2}{3}x-\frac{\alpha_{\ast}^{(1-\gamma)}}
{15\gamma}\bigg(\alpha_{\ast}-\frac{2}{3}\bigg)
\bigg(n-\frac{2}{3}\bigg)x^{3}+\cdots\ ,
\]
\begin{equation}\label{pre5}
\alpha=\alpha_{\ast}-\frac{\alpha_{\ast}^{(2-\gamma)}}{6\gamma}
\bigg(\alpha_{\ast}-\frac{2}{3}\bigg)x^{2}+\cdots\ ,
\end{equation}
where $\alpha^{*}$ is an constant of integration. The
isothermal counterpart of this solution was first
obtained by \cite{hunter1977} with $n=1$ and $\gamma=1$.

\begin{figure}
\includegraphics[scale=0.45]{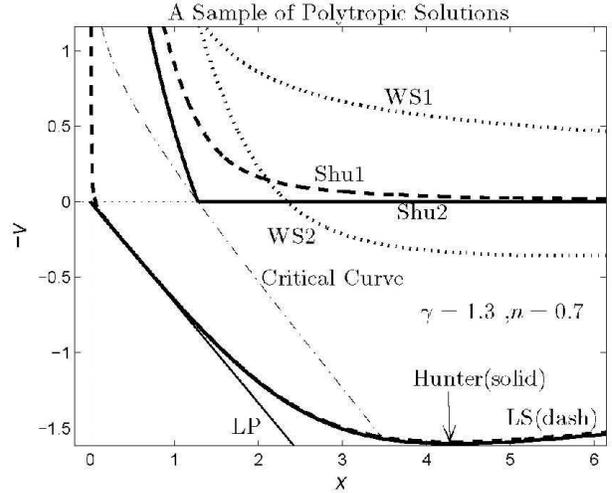}
\caption{Polytropic hydrodynamic similarity solutions with $\gamma=1.3$
and $n=0.7$ as examples of illustration for the counterparts of the
isothermal similarity solutions formerly constructed by various authors.
The dash-dotted curve represents the sonic critical curve as defined in
the text. The heavy dashed curve marked with `Shu1' is a solution with
a contraction at large $x$ and a free-fall collapse at small $x$ without
crossing the sonic critical curve. The heavy solid curve marked with
`Shu2' is the polytropic EWCS. The two heavy dotted curves labelled by
`WS1' and `WS2' are inflow and outflow solutions at large $x$ with
free-fall collapse at small $x$, respectively. The light solid solution
labelled `LP' represents a Larson-Penston type of polytropic solution.
The heavy solid curve labelled by `Hunter' is a Hunter type polytropic
solution obtained by a phase matching procedure with the sonic critical
point parameters $(\alpha,\ x,\ v)=(0.1641,\ 3.4865,\ 1.5711)$ and a
central reduced density parameter $\alpha_\ast=22.038$, and the heavy
dashed curve labelled by `LS' is a polytopic EECC solution first
constructed by Lou \& Shen (2004) for the isothermal case with
two sets of relevant sonic critical point parameters
$(\alpha,\ x,\ v)=(164.69,\ 0.002978,\ -2.4497)$ and
$(\alpha,\ x,\ v)=(0.1639,\ 3.4684,\ 1.5586)$. The last two
curves happen to be fairly close to each other when $x\gsim 0.1$.
%{\bf Other parameters for these solutions?}
%{\bf Please change the title of Figure 1
%to A Sample of Polytropic Solutions.}
}\label{PolySphereCounterparts}
\end{figure}

These solutions are summarized here
%included in \cite{sutosilk88}
as the polytropic counterparts of the relevant results in the
isothermal case. Figure \ref{PolySphereCounterparts} is a
collection of examples for previously known solutions with
parameters $\gamma=1.3$ and $n=0.7$. The EWCS constructed here
is the polytropic counterpart of the isothermal EWCS obtained
by \cite{shu1977}. Given the more general equation of state
studied by \cite{cheng78}, the isothermal EWCS represents only
a subset of all possible EWCSs (see also McLaughlin \& Pudritz
1997 in the logotropic approximation).

\begin{table}
\center\caption{Relevant parameters for the known types of
global similarity polytropic solutions constructed for $n=0.7$
and $\gamma=1.3$ (see Figure \ref{PolySphereCounterparts}) are
summarized below. Parameters $m(0)$ and $\alpha_\ast$ are
provided when the inner asymptotic behaviours are in the
free-fall type or the Hunter type,
respectively.}\label{counterpartparameter}
\begin{tabular}{ccccccc}\hline
type&$m(0)$&$A$&$B$&$\alpha_\ast$\\
\hline
Shu1&0.350&0.5&0&/\\
Shu2&0.273&0.5&$-1$&/\\
WS1&0.374&1&1&/\\
WS2&0.700&0.404413&0&/\\
LS&3.58$\times10^{-3}$&2.34&3.78&/\\
Hunter&/&2.37&3.82&22.038\\
\hline
\end{tabular}
\end{table}

\section[]{The Sonic Critical Curve}

The points where the denominators on the right-hand sides (RHSs)
of both ODEs (\ref{ode2}) and (\ref{ode3}) vanish constitute the
singular surface. On this surface, the travel speed of disturbances
relative to the local flow speed is equal to the local sound speed.
Global solutions of the two ODEs (\ref{ode2}) and (\ref{ode3})
cannot cross this singular surface smoothly, unless the intersection
happens to (i) lie on the so-called sonic critical curve, on which
both the numerators and denominators of the RHSs in the two ODEs
vanish, and (ii) have the first derivatives of $v$ and $\alpha$ with
respect to $x$ satisfying the critical conditions at this intersection
point (see Whitworth \& Summers 1985 for the isothermal case). There
are possibilities to
go across the sonic critical curve with weak discontinuities (e.g.,
Whitworth \& Summers 1985) or with shocks (Tsai \& Hsu 1985; Shu et al.
2002; Bian \& Lou 2005; Yu, Lou, Bian \& Wu 2006; Lou \& Gao 2006). The
sonic critical curve and the critical conditions are derived below
(see also McLaughlin \& Pudritz 1997).

\subsection[]{Determination of the Sonic Critical Curve}

For an isothermal gas, the sonic critical curve can be
expressed in the form of $v$ and $\alpha$ as functions
of $x$ \citep{shu1977,loushen2004}. For a polytropic gas,
the sonic critical curve can be determined numerically
for $x$ and $v$ values from specified $\alpha$ values. As
only two of the three vanishing equations of numerators
and denominators are independent,
%(i.e. substituting one equation into
%another, will yield the third one),
we just need to consider the following two equations.
\begin{equation}\label{critical1}
(nx-v)^2=\gamma\alpha^{\gamma-1}\
\end{equation}
and
\begin{equation}\label{critical2}
(n-1)v+\alpha\frac{(nx-v)}{(3n-2)}
-\frac{2(x-v)(nx-v)}{x}=0\ .
\end{equation}
For $nx-v>0$ and thus $m>0$, we then obtain
\begin{equation}\label{critical3}
v=nx-\sqrt{\gamma}\alpha^{(\gamma-1)/2}\
\end{equation}
and
\begin{equation}\label{critical4}
\bigg(n-1+\frac{\alpha}{3n-2}\bigg)
\sqrt{\gamma}\alpha^{(\gamma-1)/2}
=\frac{2\gamma}{x}\alpha^{\gamma-1}-n(n-1)x\
\end{equation}
to determine the sonic critical curve.
%The latter equation (\ref{critical4}) immediately
%leads to: \begin{equation}\label{critical5}
%n(1-n)x^2-\!\bigg[n-1+\frac{\alpha}{(3n-2)}\bigg]\!
%\sqrt{\gamma}\alpha^{(\gamma-1)/2}\!x+2\gamma\alpha^{\gamma-1}=0\ .
%\end{equation}
Given a specific $\alpha$ value (note that $\alpha>0$
by definition), the corresponding $x$ values can
be solved from equation (\ref{critical4}) as
\[
x\!=\!\!\bigg[n-1+\frac{\alpha}{3n-2}\pm\!\!
\sqrt{\bigg(n-1+\frac{\alpha}{3n-2}\bigg)^2-8n(1-n)}\bigg]
\]
\begin{equation}\label{critical6}
\qquad\times\big[2n(1-n)\big]^{-1}
\sqrt{\gamma}\alpha^{(\gamma-1)/2}\ .
\end{equation}
We then obtain the corresponding $v$ from equation
(\ref{critical3}) once $\alpha$ and $x$ are known.

\subsection[]{Completeness and Asymptotic Behaviours\\
\qquad of the Sonic Critical Curve}

According to equation (\ref{critical4}), in order
to achieve a positive $x$ value, $\alpha$ must be
larger than a critical value $\alpha_c$ defined by
\begin{equation}\label{critical7}
\alpha_c\equiv (3n-2)\big\{2\big[2n(1-n)\big]^{1/2}+1-n\big\}\ .
\end{equation}
When $\alpha>\alpha_c$, equation (\ref{critical4}) gives two
different positive $x$ roots. It is also routine to verify
that when taking the `$+$'  and `$-$' signs in equation
(\ref{critical6}), $x$ monotonically increases and decreases
with increasing $\alpha$, respectively.
%and when taking the `$-$' sign in equation (\ref{critical6}),
%$x$ monotonically decreases with $\alpha$.

The asymptotic behaviours of the sonic critical curve can
be inferred from equation (\ref{critical6}). When $\alpha$
approaches $+\infty$ and taking the `$+$' sign in equation
(\ref{critical6}), we have
\[
x\simeq\sqrt{\gamma}\alpha^{(\gamma+1)/2}\big/
\big[n(1-n)(3n-2)\big]\ ,
\]
\begin{equation}\label{critical8}
v\simeq nx\ ,
\end{equation}
while taking the `$-$' sign in equation
(\ref{critical6}), we have
\[
x\simeq2\sqrt{\gamma}(3n-2)\alpha^{(\gamma-3)/2}\ ,
\]
\begin{equation}\label{critical9}
v\simeq-\sqrt{\gamma}\alpha^{(\gamma-1)/2}\simeq\
-\sqrt{\gamma}\bigg[\frac{x}
{2\sqrt{\gamma}(3n-2)}\bigg]^{-(\gamma-1)/(3-\gamma)}\ .
\end{equation}
Equations (\ref{critical8}) and (\ref{critical9}) show
the asymptotic trends of $v$ and $x$ versus $\alpha$,
and is valuable in determining the entire sonic critical
curve. Our analysis here is complementary to the analysis
on the sonic critical points by \cite{sutosilk88}.

\begin{figure}
\includegraphics[scale=0.45]{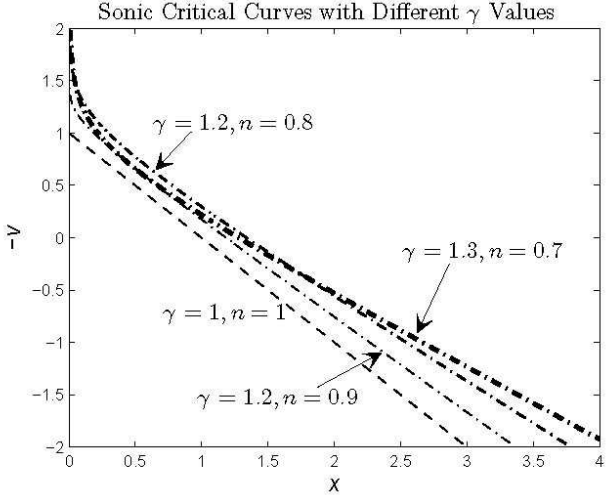}
\caption{The sonic critical curves for different pairs of $\gamma$
and $n$ values. Here, $n=2-\gamma$ is adopted for a conventional
polytropic gas. The $\gamma=n=1$ line (the dashed line) is a
straight line $v=x-1$ for the isothermal case (Shu 1977; Lou \&
Shen 2004). The major differences between isothermal and polytropic
asymptotic behaviours are the divergent behaviour at small $x$
and the different slope of $-v$ versus $x$ at large $x$ of the
polytropic cases.
}\label{PolyCritical}
\end{figure}

Fig. \ref{PolyCritical} shows the sonic critical curves
for different values of $\gamma$ and $n$ with $n=2-\gamma$.
%has been adopted in our model.
This figure shows clearly the qualitative difference in
the asymptotic behaviours between the isothermal and
polytropic cases.

\subsection[]{Eigensolutions across the Sonic Critical Curve}

As the sonic critical curve is a singular curve for the two
nonlinear ODEs (\ref{ode2}) and (\ref{ode3}), it is essential
to determine the solution behaviours in the vicinity of this
curve. \cite{sutosilk88} analyzed the solution behaviours in
the vicinity of the sonic critical curve and obtain
\[
(1+\gamma)v'^2+\bigg[n+1-4\gamma+4(\gamma-1)
\frac{v}{x}\bigg]v'
%+2(2\gamma-1)\frac{v^2}{x^2}
\]
\[
\qquad +2(2\gamma-1)\frac{v^2}{x^2}
+2\bigg[\frac{\alpha}{(3n-2)}-2n-4\gamma+4\bigg]
\frac{v}{x}
\]
\begin{equation}\label{critical10}
%\ +2\bigg(\frac{\alpha}{3n-2}-2n-4\gamma+4\bigg)\!
%\frac{v}{x}
\qquad\qquad\qquad
+\frac{(n-2)}{(3n-2)}\alpha+2(n+2\gamma-2)=0\
\end{equation}
and
\begin{equation}\label{critical11}
\alpha'=\frac{\alpha v'-2\alpha(x-v)/x}{(nx-v)}\ ,
\end{equation}
where the data set of ($x$, $v$, $\alpha$) is along the sonic critical
curve and $v'$ and $\alpha'$ are the first-order derivatives of $v$
and $\alpha$ with respect to $x$ across the sonic critical curve. In
general, there exist two eigensolutions, each corresponding to a pair
of first-order derivatives $v'$ and $\alpha'$, that can cross the sonic
critical curve. \cite{whitworthsummers1985} constructed solutions
crossing the critical curve with first-order derivatives satisfying
equations (\ref{critical10}) and (\ref{critical11}) for the isothermal
case, but with discontinuities in higher order derivatives.
%a different pair of second order derivatives with solutions
%that smoothly (analytically) cross the critical curve.
\cite{hunter1986} promptly pointed out that such solutions
involve weak discontinuities and may not be physically valid.
%by his conclusion.
In terms of global solutions, we shall mainly consider
analytically smooth solutions that cross the sonic critical curve.
%{\bf Please make the statements in this subsection more clearly.}
%{\it please see whether the above statements are clear enough.}

\section[]{New `Quasi-Static' Asymptotic Solutions}

We now describe the new `quasi-static' asymptotic
solutions in the present model framework.

\subsection[]{Singular Polytropic Sphere (SPS)
and Quasi-Static Asymptotic Solutions }

The static solution of a singular polytropic
sphere (SPS) is well known and is given by
\begin{equation}\label{static1}
v=0\ ,\qquad\qquad
\alpha=\bigg[\frac{n^2}{2\gamma(3n-2)}\bigg]^{-1/n}x^{-2/n}\ ,
\end{equation}
as a thermal-gravitational equilibrium (Suto \& Silk 1988; note
that $n=2-\gamma$ here). \cite{shu1977} used this solution to
construct the so-called expansion wave collapse solution (EWCS)
for the isothermal case with $n=1$. We realize that this solution
can in fact serve as an asymptotic `quasi-static' solution when
$x$ becomes sufficiently small. In other words, the leading terms
of $v(x)$ and $\alpha(x)$ are described by equation (\ref{static1}),
yet there may exist higher orders terms to form an asymptotic
series solution.

To be consistent, we assume up to the second orders
\begin{equation}\label{static2}
v=Lx^{K}=o(x)+\cdots\ ,
\end{equation}
\[
\alpha=\bigg[\frac{n^{2}}
{2\gamma(3n-2)}\bigg]^{-1/n}x^{-2/n}+\Delta\alpha+\cdots\ ,
\]
where $L$ and $K$ above, and $N$ below are three constants,
\begin{equation}\label{static3}
\Delta\alpha\equiv Nx^{K-1-2/n}=o(\alpha)\ ,
\end{equation}
%{\bf the precise meaning of $o(\alpha)$ in this context? }
%{\it I think it means that $\Delta\alpha/\alpha\rightarrow0$
%when $x\rightarrow0$}
with the notation $o(\alpha)$ indicating
$\Delta\alpha/\alpha\rightarrow0$ as $x\rightarrow 0^+$.
%$K$ is the corresponding index of variable $x$ for $v$.
Here, $K$ may be complex in general (see the analysis below) and
$Re(K)>1$ is required such that the additional terms in equations
(\ref{static2}) and (\ref{static3}) are indeed higher order
terms\footnote{Note that in solution (\ref{static1}), the
first-order term of $v$ is viewed to be $0x$ as a linear
function of $x$.} as compared to solution (\ref{static1}).

Substituting these expressions into equations (\ref{ode2})
and (\ref{ode3}) for small $x$, we obtain respectively
\[
\frac{n^2(K+1)N}{2}=\bigg[\frac{n^2}
{2\gamma(3n-2)}\bigg]^{-1/n}L\ ,
\]
\begin{equation}\label{static4}
n(K-1)N=\bigg(K+2-\frac{2}{n}\bigg)
\bigg[\frac{n^2}{2\gamma(3n-2)}\bigg]^{-1/n}L\ ,
\end{equation}
leading to
\begin{equation}\label{static5}
K^2-\big({4}/{n}-3\big)K+2=0\
\end{equation}
and
\begin{equation}\label{static6}
N=\frac{2}{n^2(1+K)}\bigg[\frac{n^2}
{2\gamma(3n-2)}\bigg]^{-1/n}L\ .
\end{equation}
Equation (\ref{static6}) shows that once a $K$ root is determined,
the ratio $N/L$ is then obtained. Of course, $L$ and $N$ can also
be both zero, corresponding to the static SPS solution
(\ref{static1}).

The two $K$ roots of quadratic equation (\ref{static5}) are
\begin{equation}\label{static7}
K=\frac{2}{n}-\frac{3}{2}\pm
\frac{1}{2}\bigg(1-\frac{24}{n}+\frac{16}{n^2}\bigg)^{1/2}\ .
\end{equation}
For $Re(K)>1$, $n<0.8$ is required. When $n\leq12-8\sqrt{2}$, $K$
is a real number, while for $12-8\sqrt{2}<n<0.8$, $K$ becomes a
complex number.

\subsection[]{Quasi-Static Solutions of the\\
\qquad\ First Kind with a Real $K$}

The `quasi-static' solution with a real $K$ is apparent and is
confined to a narrow range of $n$ values, i.e.,
$n\leq12-8\sqrt{2}$. There exist two sets of such solutions
corresponding to the two $K$ roots of equation (\ref{static7})
being both larger than 1.
%These two solutions can be utilized to
%construct asymptotic solutions.
In such cases, the two asymptotic solution are of the form
\[
v=Lx^K\ ,
\]
\[
\alpha=\bigg[\frac{n^{2}} {2\gamma(3n-2)}\bigg]^{-1/n}x^{-2/n}
\]
\begin{equation}\label{static8}
\qquad\qquad +\frac{2L}{n^2(1+K)}\bigg[\frac{n^2}
{2\gamma(3n-2)}\bigg]^{-1/n}x^{K-1-2/n}\ ,
\end{equation}
where the two $K$ roots are given by expression (\ref{static7})
and $L$ is an arbitrary parameter.

\subsection[]{Quasi-Static Solutions of the\\
\qquad\ Second Kind with a Complex $K$}

The `quasi-static' solution with a complex $K$ (i.e., when
$12-8\sqrt{2}<n<0.8$) appears special. For a complex $K=K_1+iK_2$
with $K_1$ and $K_2$ both being real, we should take the real
parts of both sides of equations (\ref{static2}) and
(\ref{static3}) for real $v(x)$ and $\alpha(x)$.
%but by definition $v$ and $\alpha$ are real, thus
%we should pick the real part of both sides in
%equations (\ref{static2}) and (\ref{static3}).
Then the $v(x)$ solution appears as
\[
v=Re(Lx^K)=Re\big[Lx^{K_1}\exp(iK_2\ln x)\big]
\]
\begin{equation}\label{complex1}
\qquad=x^{K_1}\big[L_1\cos(K_2\ln x)-L_2\sin(K_2\ln x)\big]\ ,
\end{equation}
where $L_1$ and $L_2$ are real and imaginary parts of parameter
$L$, respectively. By equation (\ref{static7}), we have
$K_1=2/n-3/2$ and $K_2=(-1/4+6/n-4/n^2)^{1/2}$ respectively. The
corresponding second-order term $\Delta\alpha$ is
\[
\Delta\alpha=\frac{2}{n^2}\bigg[\frac{n^2}
{2\gamma(3n-2)}\bigg]^{-1/n}\big[(1+K_1)^2+K_2^2\big]^{-1}x^{-5/2}
\]
\[
\qquad\quad\times\bigg\{\big[L_1(1+K_1)+L_2K_2\big]\cos(K_2\ln x)
\]
\begin{equation}\label{complex2}
\qquad\qquad\qquad
-\big[L_2(1+K_1)-L_1K_2\big]\sin(K_2\ln x)\bigg\}\ .
\end{equation}
Since the relative phase factor in both expressions (\ref{complex1})
and (\ref{complex2}) can be adjusted by choosing the two free
parameters $L_1$ and $L_2$, there is no loss of generality to
adopt a positive value of $K_2$.
%and the real part of equations (\ref{static2})
%and (\ref{static3}). {\bf Precise meaning?}

\section[]{Self-Similar Shocks and Polytropic Jump Conditions}

Shocks form when faster flows catch up slower flows. In our
model, it is of considerable interest to discuss self-similar
shocks with the shock front `fixed' in the self-similar profile.
Across such a shock front, the upstream and downstream regions
experience a change in the specific entropy leading to a
change in $k$ parameter in similarity transformation
(\ref{transform1}) and (\ref{transform2}).

\subsection[]{Jump Conditions for Polytropic Shocks}

We denote the upstream (i.e., a fluid flows from this side into a
shock front) physical variables with subscript 1, and the
downstream (i.e., a fluid flows on this side away from a shock
front) physical variables with subscript 2. The location of a
shock front is denoted with a subscript $s$
\begin{equation}\label{jump1}
r_s=k^{1/2}t^nx_s\ ,
\end{equation}
where $k$ is for the upstream region, i.e. $k_1=k$ or
$\kappa_1=\kappa$. For $\kappa_2=\kappa\lambda^2$ or
$k_2=k\lambda^2$ with $\lambda^2$ being a scaling parameter, the
similarity transformation in the downstream side is then
\[
\quad r_2\equiv a_2x_2\ , \qquad  u_2\equiv b_2v_2\ , \qquad
\rho_2\equiv c_2\alpha_2\ ,
\]
\begin{equation}\label{jump2}
\quad p_2\equiv d_2\beta_2\ , \qquad  M_2\equiv e_2m_2\ ,
\end{equation}
where the five time-dependent scaling functions $a_2$ to $e_2$ are
defined explicitly by
\[
\quad a_2\equiv\lambda k^{1/2}t^{n}\ ,\qquad b_2\equiv\lambda
k^{1/2}t^{n-1}\ ,\qquad c_2\equiv\frac{1}{4\pi Gt^{2}}\ ,
\]
\begin{equation}\label{jump3}
\quad d_2\equiv\lambda^2\frac{kt^{2n-4}}{4\pi G}\ ,\qquad
e_2\equiv\lambda^3\frac{k^{3/2}t^{3n-2}}{(3n-2)G}\ .
\end{equation}
At the shock front, we should
have $r_1=r_2=r_s$ and thus
\begin{equation}\label{jump4}
x_1=x_s=\lambda x_2\ .
\end{equation}
Given $x_1,\ v_1$ and $\alpha_1$, we need only to calculate the
corresponding $x_2,\ v_2$ and $\alpha_2$ across a shock, and
the parameter $\lambda$ is automatically determined by equation
(\ref{jump4}) and hence the downstream transformation
(\ref{jump2}) and (\ref{jump3}) is known.

We now describe the jump conditions for shocks in a polytropic
gas. Using the standard procedure, we choose the framework of
reference in which the shock front is instantly at rest. The
jump conditions in this shock framework of reference are then
given below. The mass conservation is
\begin{equation}\label{jump5}
\big[\rho(u_s-u)\big]_1^2=0\ ,
\end{equation}
where $u_s\equiv\mbox{d}r_s/\mbox{d}t$ is the radially outward
travel speed of the shock front and, following the notational
convention, we denote the difference of the corresponding
physical variables between the upstream and downstream sides
by enclosing them within a pair of square brackets with a
superscript 2 and a subscript 1. The momentum conservation is
\begin{equation}\label{jump6}
\left[\kappa\rho^\gamma+\rho(u_s-u)^2\right]_1^2=0\ .
\end{equation}
The energy conservation becomes
\begin{equation}\label{jump7}
\left[\frac{\rho(u_s-u)^3}{2} +\frac{\gamma\kappa
(u_s-u)}{(\gamma-1)}\rho^{\gamma}\right]_1^2=0\ .
\end{equation}

Combining equations (\ref{jump5}), (\ref{jump6}) and (\ref{jump7})
together with the self-similar transformation, we derive three
shock conditions in terms of the self-similar variables
\begin{equation}\label{jump8}
\alpha_1(nx_1-v_1)=\lambda\alpha_2(nx_2-v_2)\ ,
\end{equation}
\begin{equation}\label{jump9}
\alpha_1^{\gamma}+\alpha_1(nx_1-v_1)^2=\lambda^2
\left[\alpha_2^{\gamma}+\alpha_2(nx_2-v_2)^2\right]\ ,
\end{equation}
%and
\[
(nx_1-v_1)^2+\frac{2\gamma}{(\gamma-1)}\alpha_1^{\gamma-1}
\]
\begin{equation}\label{jump10}
\qquad\qquad=\lambda^2\left[(nx_2-v_2)^2+\frac{2\gamma}
{(\gamma-1)}\alpha_2^{\gamma-1}\right]\ .
\end{equation}
\begin{figure}
\includegraphics[scale=0.45]{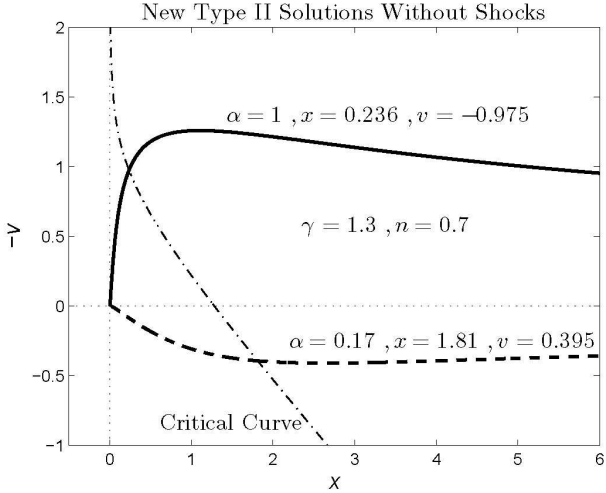}
\caption{Semi-complete similarity solutions approaching type II
`quasi-static' asymptotic solutions at small $x$ without shocks
for $\gamma=1.3$ and $n=0.7$. Here, $-v$ versus $x$ is shown.
The dash-dotted line is the sonic critical curve. The solid
solution is an inflow solution which is obtained by integrating
from $(x,\ v,\ \alpha)=(0.236,\ -0.975,\ 1)$ on the critical
curve inward; and the dashed curve is an outflow solution
which is obtained by integrating from $(x,\ v,\ \alpha)=
(1.81,\ 0.395,\ 0.17)$ on the critical curve inward. The two
dotted straight lines are the abscissa and ordinate axes,
respectively. For a sufficiently small $x$, both solutions
approach the Type II `quasi-static' asymptotic solution.
%{\bf Please change the title to
%New Type II Solutions without Shocks.}
}\label{PolyWithout1}
\end{figure}
\begin{figure}
\includegraphics[scale=0.45]{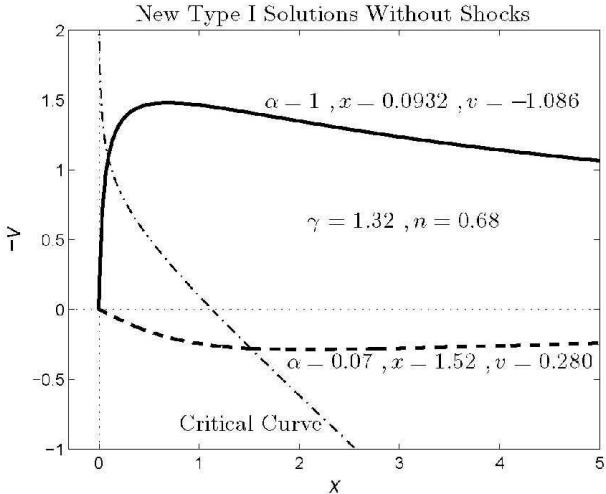}
\caption{Semi-complete similarity solutions approaching type I
`quasi-static' asymptotic solutions at small $x$ without shocks
for $\gamma=1.32$ and $n=0.68$. Here, $-v$ versus $x$ is shown.
The dash-dotted line is the sonic critical curve. The solid curve
is an inflow solution which is obtained by integrating from
$(x,\ v,\ \alpha)=(0.0932,\ -1.086,\ 1)$ on the sonic critical
curve inward; and the dashed curve is an outflow solution obtained
by integrating from $(x,\ v,\ \alpha)=(1.52,\ 0.280,\ 0.07)$ on
the sonic critical curve inward. The two dotted straight lines are
the abscissa and ordinate axes, respectively. For a sufficiently
small $x$, both solutions approach the Type I `quasi-static'
asymptotic solution.
%{\bf Please change the title to New Type I Solutions without Shocks.}
}\label{PolyWithout2}
\end{figure}

\subsection[]{Shock Solutions by the Jump Conditions}

By introducing new variables $\Gamma_i\equiv n-x_i/v_i\ $
with $i=1,\ 2$, we reduce equations (\ref{jump8}) through
(\ref{jump10}) to
\begin{equation}\label{sol1}
\alpha_1\Gamma_1=\alpha_2\Gamma_2\ ,
\end{equation}
\begin{equation}\label{sol2}
\frac{\alpha_1^{\gamma}}{x_1^2}+\alpha_1\Gamma_1^2
=\frac{\alpha_2^{\gamma}}{x_2^2}+\alpha_2\Gamma_2^2\ ,
\end{equation}
\begin{equation}\label{sol3}
\Gamma_1^2+\frac{2\gamma}{(\gamma-1)}\frac{\alpha_1^{\gamma-1}}{x_1^2}
=\Gamma_2^2+\frac{2\gamma}{(\gamma-1)}
\frac{\alpha_2^{\gamma-1}}{x_2^2}\ .
\end{equation}
We intend to solve for $\Gamma_2$, $\alpha_2$ and $x_2$ with known
values of $\Gamma_1$, $\alpha_1$ and $x_1$. By substituting
$\alpha_2=\alpha_1\Gamma_1/\Gamma_2$ of equation (\ref{sol1}),
solving for $x_2^2$ from equations (\ref{sol2}) and (\ref{sol3})
respectively, and eliminating $x_2^2$ accordingly, we arrive at
the following quadratic equation in terms of $\Gamma_2$
\[
\frac{(\gamma+1)}{2\gamma}\Gamma_2^2
-\frac{(\alpha_1^{\gamma}/x_1^2+\alpha_1\Gamma_1^2)}
{\alpha_1\Gamma_1}\Gamma_2
\]
\begin{equation}\label{sol4}
\qquad\qquad\qquad
+\frac{(\gamma-1)}{2\gamma}\left(\Gamma_1^2+\frac{2\gamma}{\gamma-1}
\frac{\alpha_1^{\gamma-1}}{x_1^2}\right)=0\ .
\end{equation}
Excluding the trivial solution $\Gamma_2=\Gamma_1$, we
obtain one root of equation (\ref{sol4}) in the form of
\begin{equation}\label{sol5}
\Gamma_2=\frac{2\gamma
\alpha_1^{\gamma-1}}{(\gamma+1)x_1^2\Gamma_1}
%\frac{\alpha_1^{\gamma-1}}{x_1^2\Gamma_1}
+\frac{(\gamma-1)}{(\gamma+1)}\Gamma_1\ .
\end{equation}
In terms of the upstream Mach
number ${\cal M}_1$ defined by
\begin{equation}\label{sol6}
{\cal M}_{1}^2\equiv
\frac{(u_s-u_1)^2}{s_1^2}=\frac{\rho_1(u_s-u_1)^2}{\gamma
p_1}=\frac{x_1^2\Gamma_1^2}{\gamma\alpha_1^{\gamma-1}}\ ,
\end{equation}
expression (\ref{sol5}) reads
\begin{equation}\label{sol7}
\frac{u_s-u_2}{u_s-u_1}=\frac{\Gamma_2}{\Gamma_1}=\frac{2}
{(\gamma+1){\cal M}_{1}^2}+\frac{(\gamma-1)}{(\gamma+1)}\ ;
\end{equation}
this is precisely equation (89.6) of \cite{landau1959}.

\begin{figure}
\includegraphics[scale=0.45]{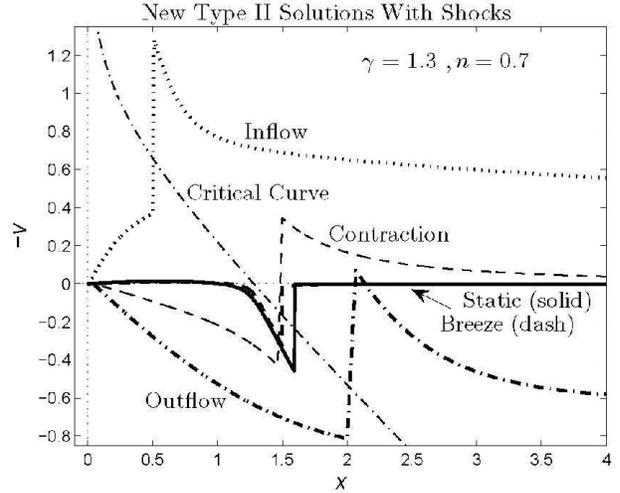}
\caption{Semi-complete similarity shock solutions containing Type II
`quasi-static' asymptotic solutions for $\gamma=1.3$ and $n=0.7$ are
shown here in terms of $-v$ versus $x$. The heavy dotted curve is a
shock solution with an inflow in the outer portion, constructed by
integrating from $(x,\ v,\ \alpha)=(0.751,\ -0.426,\ 0.3)$ on the
sonic critical curve with $x_{s_2}=0.5$; the heavy dash-dotted line
is a solution with an outflow in the outer portion, constructed by
integrating from $(x,\ v,\ \alpha)=(2.47,\ 0.866,\ 0.16)$ on the sonic
critical curve with $x_{s_2}=2$; the light dashed curve is a solution
with a contracting outer portion, constructed with parameters $A=0.5$,
$B=0$ and a chosen $x_{s_1}=1.5$; the heavy solid curve is a solution
with a static outer portion (i.e., part of a SPS), constructed with
parameters $A=0.404$, $B=0$ and a chosen $x_{s_1}=1.6$; the heavy
dashed curve is a shock solution with a breeze in the outer portion,
constructed with parameters $A=0.402$ and $B=0$ and choosing
$x_{s_1}=1.5$. The two light dotted straight lines are the abscissa
and ordinate axes, respectively. For small $x$, all solutions
approach the Type II `quasi-static' asymptotic solution.
%{\bf Please change the word `Contract' to `Contraction'
%and change the title to New Type II Solutions with Shocks.}
}\label{PolyWith1}
\end{figure}

\begin{figure}
\includegraphics[scale=0.45]{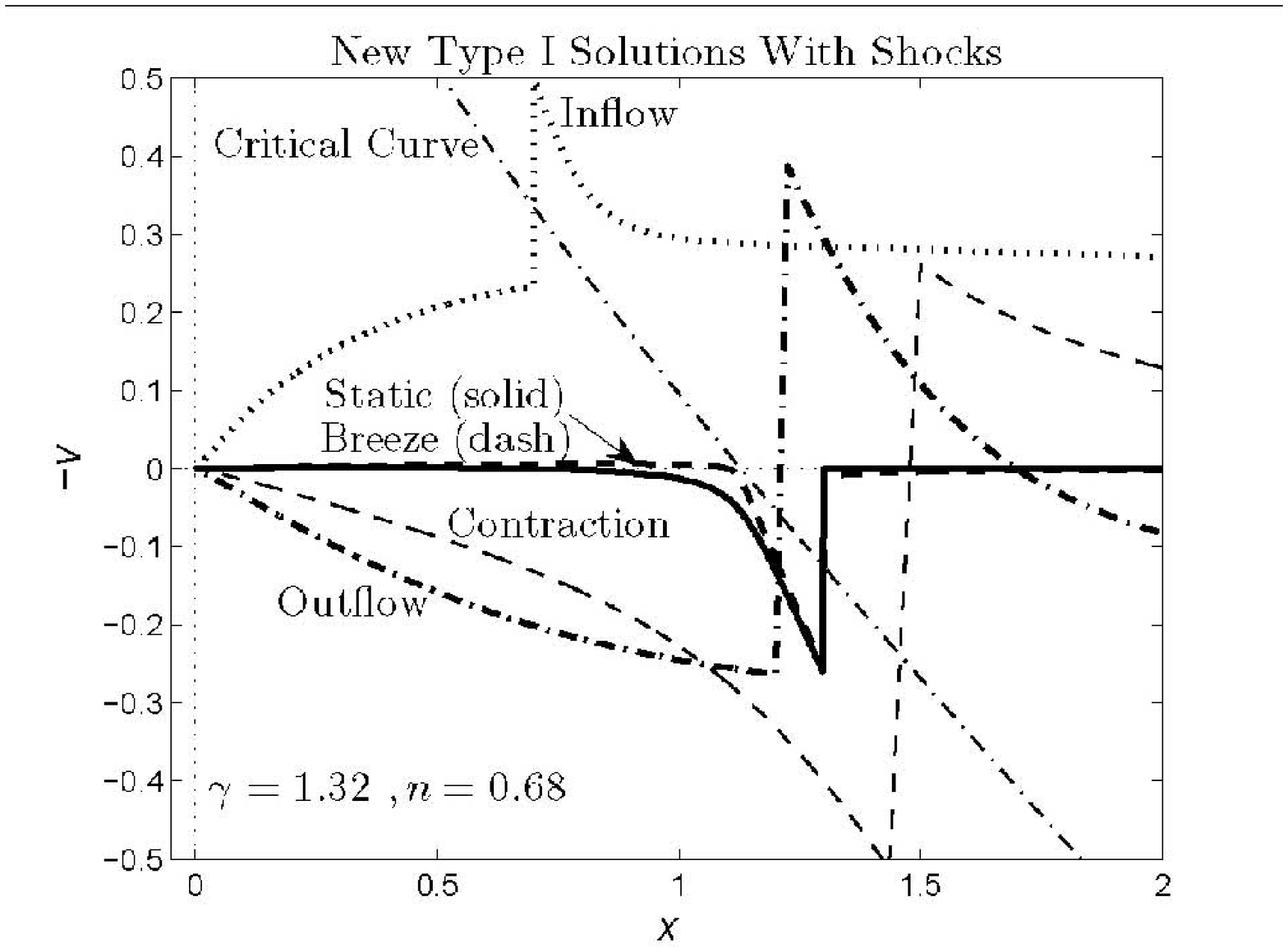}
\caption{Semi-complete similarity shock solutions containing Type I
`quasi-static' asymptotic solutions for $\gamma=1.32$ and $n=0.68$ are
shown in terms of $-v$ versus $x$. The heavy dotted line is a solution
with an inflow in the outer portion, constructed by integrating from
$(x,\ v,\ \alpha)=(0.812,\ -0.243,\ 0.1)$ on the sonic critical curve
with a chosen $x_{s_2}=0.7$; the heavy dash-dotted line is a solution
with an outflow in the outer portion and constructed by integrating
from $(x,\ v,\ \alpha)=(1.52,\ 0.280,\ 0.07)$ on the sonic critical
curve with a chosen $x_{s_2}=1.2$; the light dashed curve is a solution
with a contracting envelope, constructed with parameters $A=0.15$ and
$B=0$ and a chosen $x_{s_1}=1.5$; the heavy solid curve is a solution
with a static outer portion of SPS, constructed with parameters
$A=0.114$ and $B=0$ and a chosen $x_{s_1}=1.3$; the heavy dashed
curve is a solution with a breeze in the outer portion, constructed
with parameters $A=0.1135$ and $B=0$ and a chosen $x_{s_1}=1.3$.
The two dotted straight lines are the abscissa and ordinate axes,
respectively. All solutions approach the Type I `quasi-static'
asymptotic solution at small $x$.
%{\bf Please change the word `Contract' to `Contraction'
%and change the title to New Type I Solutions with Shocks.}
}\label{PolyWith2}
\end{figure}

It is straightforward to prove that the following set of
inequalities are compatible with each other (${\cal M}_2$
is the downstream Mach number)
\[
\frac{\Gamma_2}{\Gamma_1}<1\ ,\qquad
\frac{u_s-u_2}{u_s-u_1}<1\ ,
\qquad {\cal M}_1^2>1\ ,\qquad {\cal M}_2^2<1\ ,
\]
\begin{equation}\label{sol8}
\lambda>1\ ,\quad\qquad k_2>k_1\ ,
\qquad\qquad \kappa_2>\kappa_1\ .
\end{equation}
We only give a proof for the compatibility between $\lambda>1$
and ${\cal M}_1^2>1$ below as an example of illustration. First note
that when ${\cal M}_1^2>1$, we have $z\equiv\Gamma_2/\Gamma_1<1$. From
equation (\ref{sol2}), we then solve for $\lambda^2=x_1^2/x_2^2$ as
\begin{equation}\label{sol9}
\lambda^2=z^\gamma\big[\gamma
{\cal M}_1^2(1-z)+1\big]=\frac{2\gamma(1-z)z^\gamma}
{(\gamma+1)z-(\gamma-1)}+z^\gamma\ .
\end{equation}
The first derivative of $\lambda^2$ with respect to $z$ gives
\begin{equation}\label{sol10}
\frac{\mbox{d}\lambda^2}{\mbox{d}z}=-\frac{\gamma(\gamma^2-1)
(z-1)^2z^{\gamma-1}}{\big[(\gamma+1)z-(\gamma-1)\big]^2}<0\
\end{equation}
for $\gamma>1$. Since $\lambda^2\rightarrow1$ as $z\rightarrow1$,
it follows that when $z<1$ (i.e., ${\cal M}_1^2>1$), $\lambda>1$
holds and vice versa. According to the Zempl\'en theorem [see, e.g.,
\cite{landau1959}], for entropy to increase across a shock going
from the upstream to downstream side, we have physical shock
solutions with $u_s-u_2<u_s-u_1$. While this general conclusion
only holds for weak shocks, it is true in our case from equivalent
inequalities (\ref{sol8}). That is, the upstream and downstream
regions should be supersonic and subsonic, respectively.

\subsection[]{Solutions and Physical Interpretation}

%For the convenience of analysis, we adopt the following notations.
%%Hereafter we change our notations as below for convenience:
%We do not distinguish between $x_1$ and $x_2$ in the unshocked
%region, and use $x_{s_1}$ and $x_{s_2}$ for the shock location
%in the similarity transformations of the upstream and downstream
%regions, respectively. The ratio $\lambda=x_{s_1}/x_{s_2}$ is
%then fixed.
%{\it
For the convenience of analysis and comprehension, we adopt
the solution procedure and presentation in the following manner.
We do not distinguish between $x_1$ and $x_2$ in the unshocked
regions, even though the self-similar transformations are
different for the upstream and downtream regions. We use
$x_{s_1}$ and $x_{s_2}$ for the shock location under the similarity
transformations of the upstream and downstream sides, respectively.
The ratio $\lambda=x_{s_1}/x_{s_2}$ is then fixed by shock jump
conditions. In figure illustrations of shock solutions, we juxtapose
curves for $-v_1$ versus $x_1$ and for $-v_2$ versus $x_2$ in the
same figure, with the implied differences in downstream and
upstream transformations, and link the two shock points
$(-v_{s_1},\ x_{s_1})$ and $(-v_{s_2},\ x_{s_2})$ by a straight
line, indicating that these two points represent the demarcation of
upstream and downstream quantities on the two sides of the shock
front, respectively. A reader of these figures may interpret the
curves to the left of this straight line ($x<x_{s_2}$) as inner
downstream solutions, and the curves to the right of this straight
line ($x>x_{s_1}$) as the corresponding outer upstream solutions.
As several shock solutions are pact together within one figure,
one should also note that for different shock solutions, this
demarcation is also different because $x_{s}$ values are different.
The advantage of such presentations is that the critical curve of
both regions coincide in the figure, i.e., the curve provided in
these figures serves as the critical curve in both regions.
Furthermore, one can obtain the $x_{s_1}$ and $x_{s_2}$ values
directly from the figure to compute $\lambda$ values accordingly.
In our figures, the scalings of physical quantities are different
in the presentation for upstream and downstream regions, and the
scalings are only for the reduced quantities. For example, one unit
in the two regions of the figure represents the same difference of
$x$ values, but not the same difference of $r$ values; the situation
is similar in terms of the correspondence between $v$ and $u$ values.
%}

Once we obtain a shock solution as described above, the procedure
of obtaining a physical solution of our hydrodynamic model is to
assign a $k$ value of either region, i.e., $k_1$ or $k_2$, and
obtain the $k$ value for the other region by the simple relation
$k_2=k_1\lambda^2$. With the separate self-similar transformations
in the upstream and downstream regions, we readily compute the
corresponding physical quantities of the solution. In the
dimensional form, the difference in the scaling of physical
quantities due to the similarity transformations disappears. For
example, $r_{s_1}=r_{s_2}$ now, and a unit in the two regions in
the figures represents the same difference of $r$ values.

\section[]{Semi-Complete Solutions with\\
\qquad the `Quasi-Static' Asymptotic\\
\qquad Behaviours}

Semi-complete similarity solutions include those that do not intersect
with the singular surface, that go across the sonic point analytically,
and that cross the sonic critical curve with shocks in the region
$0<x<+\infty$ (see Lou \& Shen 2004; Bian \& Lou 2005; Yu et al. 2006).
Either with or without shocks, semi-complete solutions approaching the
`quasi-static' asymptotic behaviours at small $x$ are constructed in
the present model. Specifically, we are able to construct self-similar
shock solutions approaching the `quasi-static' asymptotic behaviours
at small $x$ yet with inflow, outflow, SPS, breeze or contraction in
the outer envelope. In this section, we choose $n=0.7$ and $n=0.68$
for type II and type I `quasi-static' asymptotic behaviours at small
$x$, respectively.

\subsection[]{Smooth Similarity Solutions without Shocks}

\begin{table}
\center\caption{Parameters $A$, $B$ and $L$ of asymptotic
behaviours for solutions without shocks presented in Figures
\ref{PolyWithout1} and \ref{PolyWithout2}.}\label{asymwithout}
\begin{tabular}{ccccc}\hline
$n$&type&$A$&$B$&$L$\\ \hline
0.7 &outflow&0.673 &0.852  &$1.11+0.735i$\\
0.7 &inflow &0.0840&$-2.27$&$-11.4+19.7i$\\
0.68&outflow&0.176 &0.573  &0.452\\
0.68&inflow &0.0142&$-2.55$&$-301$\\
\hline
\end{tabular}
\end{table}
We are able to construct semi-complete similarity solutions
without shocks simply by integrating inward from a certain point
on the sonic critical curve, using the analytical eigensolution
with a smaller $v'$ in the close vicinity of this chosen point
[equations (\ref{critical10}) and (\ref{critical11}) determine the
eigensolution]. Figures \ref{PolyWithout1} and \ref{PolyWithout2}
are examples of illustration for semi-complete similarity
solutions without shocks, approaching type II and type I
`quasi-static' solutions at small $x$, respectively. In Figure
\ref{PolyWithout2}, only similarity solutions with smaller $K$
values in equation (\ref{static7}) are constructed by integrating
from the sonic critical curve using one of the eigensolutions.
%{\bf Why? How about larger $K$ root?}{\it Because the only way
%we can construct our quasi-static asymptotic solutions is to
%integrate from the critical curve using one eigensolution,
%and I have not found solutions with the larger $K$ root.}

\subsection[]{Similarity Solutions with Shocks}

\begin{table}
\center\caption{Parameters $A$, $B$ and $L$ of asymptotic
behaviours and shock parameters $x_{s_1}$, $x_{s_2}$ and $\lambda$
for global similarity shock solutions approaching type II
`quasi-static' asymptotic behaviours at small $x$ as shown in
Figure \ref{PolyWith1}.}\label{asymwith1}
\begin{tabular}{ccccccc}\hline
$\!\!\!$type$\!$&$\!A\!\!$&$\!\!B\!\!$&$\!\!L\!\!\!\!$
&$\!x_{s_2}\!$&$\!x_{s_1}\!\!$&$\!\lambda\!\!$\\
\hline $\!\!\!$inflow$\!$&$\!$0.227$\!\!$&$\!$$-1.15$$\!\!$
&$\!\!$$2.09-0.841i$$\!\!\!\!$&$\!$0.5$\!$&$\!$0.511$\!\!$&$\!$1.022$\!\!$\\
$\!\!\!$outflow$\!$&$\!$1.19$\!\!$&$\!$1.56$\!\!$
&$\!\!$3.57+1.87$i\!\!\!\!$&$\!$2$\!$&$\!$2.07$\!$&$\!\!$1.034$\!\!$\\
$\!\!\!$contraction$\!$&$\!$0.5$\!\!$&$\!$0$\!\!$
&$\!\!$0.289+0.188$i\!\!\!\!$&$\!$1.46$\!$&$\!$1.5$\!$&$\!\!$1.024$\!\!$\\
$\!\!\!$SPS$\!$&$\!$0.404$\!\!$&$\!$0$\!\!$
&$\!\!$$0.00233-0.0973$$i\!\!\!\!$&$\!$1.59$\!$&$\!$1.6$\!$&$\!\!$1.0058$\!\!$\\
$\!\!\!$breeze$\!$&$\!$0.402$\!\!$&$\!$0$\!\!$
&$\!\!$$-0.00207-0.0562$$i\!\!\!\!$&$\!$1.497$\!$
&$\!$1.5$\!$&$\!\!$1.0018$\!\!$\\
\hline
\end{tabular}
\end{table}

Based on the jump conditions, we are able to construct similarity
shock solutions with inflow, outflow, breeze, SPS or contraction
in the outer portion, connected to either type II or type I
`quasi-static' asymptotic solutions at small $x$. In constructing
semi-complete shock solutions with inflow or outflow for the outer
asymptotic behaviours, we use an inner solution that can cross the
sonic critical curve smoothly by integrating inward from the sonic
critical curve using an eigensolution, choose an $x_{s_2}$ value
in the downstream region along this inner solution, determine the
upstream quantities by the shock jump conditions, and integrate
outward from $x_{s_1}$ for the outer part of the solution. On the
other hand, in order to construct similarity solutions with
breeze, SPS or contraction in the outer portions, we integrate
inward from a large $x$ value (say $10^5$) to reach a certain
value $x_{s_1}$, determine the downstream quantities and integrate
further inward to small $x$ values. Only when the inner solutions
approach the `quasi-static' asymptotic solution do we succeed in
constructing such global shock solutions. Figures \ref{PolyWith1}
and \ref{PolyWith2} are examples of illustration for self-similar
shock solutions approaching type II and type I `quasi-static'
asymptotic behaviours at small $x$, respectively. In both Figs.
\ref{PolyWith1} and \ref{PolyWith2}, examples of inflow, outflow,
contraction, breeze and SPS for outer portions are illustrated.
For type I `quasi-static' asymptotic solutions, we again obtain
solutions only with smaller $K$ values in equation
(\ref{static7}).

\begin{table}
\center\caption{Parameters $A$, $B$ and $L$ for asymptotic
behaviours and shock parameters $x_{s_1}$, $x_{s_2}$ and $\lambda$
for similarity shock solutions with type I `quasi-static'
asymptotic behaviours shown in Fig.
\ref{PolyWith2}.}\label{asymwith2}
\begin{tabular}{ccccccc}\hline
$\!\!$type$\!\!$&$A$&$B\!$&$L\!$&$x_{s_2}$&$x_{s_1}$&$\!\!\lambda\!\!$\\
\hline
$\!\!$inflow$\!\!$&0.0765&$-0.498$$\!$&$-1.78$$\!$&0.7&0.7009&$\!\!$1.0013$\!$\\
$\!\!$outflow$\!\!$&0.184&0.412$\!$&0.452$\!$&1.2&1.225&$\!\!$1.021$\!$\\
$\!\!$contract$\!\!$&0.15&0$\!$&$-0.203$$\!$&1.43&1.5&$\!\!$1.046$\!$\\
$\!\!$static$\!\!$&0.114&0$\!$&$-0.0196$$\!$&1.298&1.3&$\!\!$1.0018$\!$\\
$\!\!$breeze$\!\!$&0.1135&0$\!$&$-0.0248$$\!$&1.298&1.3&$\!\!$1.0015$\!$\\
\hline
\end{tabular}
\end{table}

\subsection[]{Asymptotic Behaviours of Sample Solutions}

\begin{figure}
\includegraphics[scale=0.45]{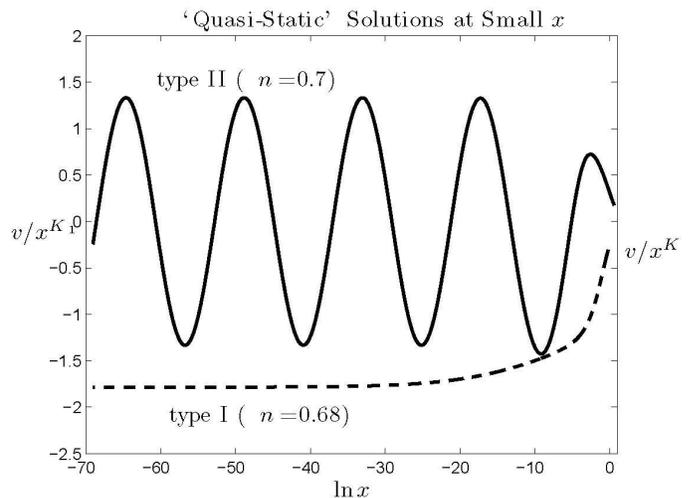}
\caption{`Quasi-static' asymptotic solutions for small $x$ values.
The presentation is in the $v/x^{K}$ versus $\ln x$ form for type I
`quasi-static' asymptotic solution, and in the $v/x^{K_1}$ versus
$\ln x$ form for type II `quasi-static' asymptotic solution.
The type I solution is obtained by integrating inward from
$(x,\ v,\ \alpha)=(0.812,\ -0.243,\ 0.1)$ on the sonic critical
curve, and the type II solution is obtained by integrating inward
from $(x,\ v,\ \alpha)=(1.81,\ 0.395,\ 0.17)$ on the sonic critical curve.
%{\bf Along the sonic critical curve? Please confirm. }{\it Yes, it is.}
%{\bf Please change the title to `Quasi-Static' Solutions at Small $x$.}
}\label{PolyAsymptote}
\end{figure}

All solutions presented in this section approach the
`quasi-static' (either type I or type II) asymptotic behaviours at
small $x$ values. Figure \ref{PolyAsymptote} displays two examples
illustrating such asymptotic behaviours in terms of $v/x^K$ versus
$\ln x$ for type I solution and $v/x^{K_1}$ versus $\ln x$ for
type II solution. For the type II 'quasi-static' solution, the sine
(cosine) pattern is apparent, while for the type I 'quasi-static'
solution, the tendency to a constant value is also fairly clear.
For the convenience of comparison and reference, we also list all
the relevant parameters $x_{s_1}$, $x_{s_2}$ and $\lambda$ for
shocks and $A$, $B$ and $L$ for asymptotic behaviours in Tables
\ref{asymwithout}, \ref{asymwith1} and \ref{asymwith2}.

%\begin{figure}
%\includegraphics[scale=0.45]{PolySample.eps}
%\caption{A sample solution for the hydrodynamic model. The
%dash-dotted curve is the sonic critical curve. This solution is
%obtained by integrating from $(x,\ v,\ \alpha)=(2.533,\ 0.9074,\
%0.1598)$ inward and choosing $x_{s_2}=1.6196$. The relevant
%parameters for asymptotic behaviours are $L=3.839+2.275i$,
%$B=-0.1087$, $A=7530$ and for shock conditions are
%$x_{s_1}=37.430$, $\lambda=23.11$. The two dotted lines are
%abscissa and ordinate axes, respectively. The vanishing point,
%defined as the largest $x$ of zero velocity, is $x_{v=0}=0.03842$.
%The straight line in the middle represents a shock, and although
%the two $x_s$ values are different, they correspond to the same
%radius $r_s$.
%%{\bf Please change the title to A Rebound Shock Solution.}
%%}
%\label{PolySample}
%\end{figure}

\begin{figure}
\includegraphics[scale=0.45]{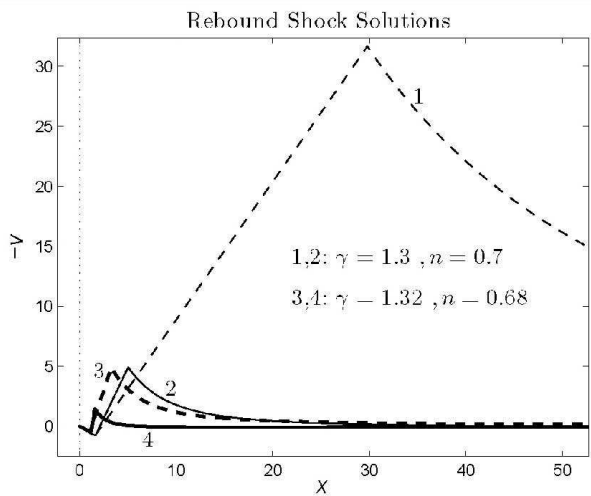}
\caption{Sample shock solutions for polytropic gas dynamics.
Solutions 1 and 2 are obtained by integrating inward from $(x,\
v,\ \alpha)=(2.533,\ 0.9074,\ 0.1598)$ on the sonic critical curve
and choosing $x_{s_2}=1.6198$ and $x_{s_2}=1.64$, respectively.
Solutions 3 and 4 are obtained by integrating inward from $(x,\
v,\ \alpha)= (1.793,\ 0.4374,\ 0.067)$ on the sonic critical curve
and choosing $x_{s_2}=1.1$ and $x_{s_2}=1.2$, respectively. The
relevant parameters for asymptotic behaviours and for shock
conditions are summarized in Table \ref{sampleparameter}. The
two dotted straight lines are abscissa and ordinate axes,
respectively. The straight lines in the middle represent shocks,
and the two $x_s$ values correspond to the same $r_s$.
%{\bf Please change the title to A Rebound Shock Solution.}
} \label{PolySample}
\end{figure}

\begin{table}
\center\caption{Parameters for asymptotic solutions and shocks in
the polytropic shock solutions displayed in Figure \ref{PolySample}.}
\label{sampleparameter}
\begin{tabular}{ccccc}\hline
No.&1&2&3&4\\
\hline
$\gamma$&1.3&1.3&1.32&1.32\\
$A$&3920&23.5&5.40&0.493\\
$B$&$-0.00442$&0.562&$-0.792$&0.371\\
$L$&$3.83+2.27i$&$3.84+2.28i$&0.441&0.441\\
$x_{s_1}$&29.79&5.013&3.288&1.578\\
$x_{s_2}$&1.6198&1.64&1.1&1.2\\
$\lambda$&18.39&3.056&2.989&1.315\\
\hline
\end{tabular}
\end{table}

\section[]{Self-Similar Rebound Shocks\\
\quad\ in Core Collapsing Stars}
\subsection[]{Physical Scenario}

The core collapse of a massive star involves a multitude of
physical processes of nuclear burning, radiation pressure,
degenerate materials, electrons, and neutrinos (Hirata et
al. 1987; Bionta et al. 1987; Bethe 1993). Here, we only
consider several gross aspects from the point of view of
fluid dynamics involving self-gravity and try to outline a
free-fall core collapse and a rebound shock scenario for a
supernova explosion.

At the onset of gravitational core collapse inside a massive star,
the central region suddenly loses pressure and a nearly free infall
towards the center ensues; meanwhile, the information of core collapse
travels outward through the stellar interior to reach the envelope
and a self-similar collapse phase grossly characterized by asymptotic
free-fall solution (\ref{pre4}) at small $x\propto r/t^n$ may
gradually emerge. By physical estimates, this happens on a timescale
of a few seconds. For small $x$, we obtain\footnote{Bethe (1993)
indicated $u\propto r^{-1/2}$ and $\rho\propto t^{-1}r^{-3/2}$
outside the (rebound) shock and after
the start of (core) collapse. This would correspond to an index $n=2/3$;
for a conventional polytropic gas with $n=2-\gamma$, this would in
turn imply $\gamma=4/3$ for an extremely relativistically hot gas
(e.g., Goldreich \& Weber 1980). We take $n>2/3$ in our model
framework to make the enclosed mass $M$ physically meaningful
in self-similar transformation (\ref{transform2}).} the radial
flow speed $u\propto -t^{(3n/2)-1}r^{-1/2}$, the mass density
$\rho\propto t^{(3n/2)-2}r^{-3/2}$, the enclosed mass
$M\rightarrow k^{3/2}m(0)t^{3n-2}/[(3n-2)G]$, and the core mass
accretion rate $\dot M\rightarrow k^{3/2}m(0)t^{3(n-1)}/G$ with
$n>2/3$. It should be noted that before the onset of core collapse,
the progenitor star is expected to have a wind in general and the
star itself can engage in global oscillations, be they acoustic
$p-$modes or internal gravity $g-$modes. For example, for purely
radial stellar acoustic oscillations, the stellar envelope and
interior can move radially either outward or inward depending on
the phase of oscillation during the rapid core collapse. These
plausible physical conditions may lead to different self-similar
evolution behaviours during a free-fall core collapse. Also, the
material degeneracy can already set in the core well before the
onset of gravitational core collapse. Moreover, the mass density
divergence associated with the free-fall collapse solution
(\ref{pre4}) can rapidly trigger core degeneracy. In reality,
no inifinity should arise.

As mass accumulates towards the center, singularity arises
mathematically and similarity disappears. As the degenerate
core pressure rapidly builds up to resist the material infall, a
powerful rebound shock emerges around the core and drives out
most gas materials as it ploughs through the infalling stellar
envelope, leaving behind a remnant compact object (such as a
neutron star or a black hole\footnote{We will also discuss
the possibility of producing or exposing a white dwarf in a
similar process separately. For a low-mass progenitor star,
the formation of a proto-white dwarf in the core may or may
not involve a violent rebound shock explosion.}) with high
mass density in the degenerate core.
%\footnote{For a low-mass progenitor star, the formation of
%a proto-white dwarf in the core may not involve a violent
%explosion.} Let us focus on neutron stars as an example of
%discussion.
For instance, for the formation of a neutron star, the breakout of
such a rebound shock through the stellar photosphere is crucial to
initiate a supernova explosion observed at terrestrial observatories
(e.g., Bethe 1995; Burrows 2000 and Lattimer \& Prakash 2004 for
reviews of
supernova and neutron star physics and extensive references therein).
Again, we propose that after the emergence of a rebound shock around
the collapsing core, a self-similar evolution grossly characterized
by either `quasi-static' asymptotic solution (\ref{static8})
or quasi-static' asymptotic solution (\ref{complex1}) and
(\ref{complex2}) may gradually appear and persist as the rebound
shock travel outward against infalling materials. The self-similar
evolution phase of a rebound shock may last on timescales in the
range of $\sim 10^4-10^5$ s or longer.

For `quasi-static' asymptotic solution
(\ref{static8}) of the first type at small $x$, we have
$u=Lk^{(1-K)/2}t^{-n(K-1)-1}r^K$, $\rho\propto r^{-2/n}
+\epsilon t^{-n(K-1)}r^{K-1-2/n}$, $M\propto r^{3-2/n}$,
and $\dot M\propto Lt^{-n(K-1)-1}r^{K+2-2/n}$, where $L<0$
for an inflow, $K>1$, and $\epsilon$ is a sufficiently
small parameter. In other words, the radial flow speed $u$
decreases with time and approaches zero as $r\rightarrow 0$;
the mass density $\rho$ diverges as $r\rightarrow 0$ to
induce material degeneracy; being
independent of time $t$, the enclosed mass $M$ vanishes as
$r\rightarrow 0$; and the core mass accretion rate $\dot M$
decreases with time and approaches zero as $r\rightarrow 0$
by quadratic equation (\ref{static7}). A very important and
testable prediction of this first type of self-similar
rebound shock explosions is that the mass density $\rho$ for
stellar materials scales as $r^{-2/n}$ with $2/n<3$ behind
the rebound shock after a sufficiently long lapse of time;
for a conventional polytropic gas with $n=2-\gamma$ and
$\gamma\geq 1$, we would also have $2<2/n$. Ideally, this
mass density profile can be better preserved with a more or
less quiet central nonrotating neutron star or Schwarzschild
black hole, otherwise central activities or winds inevitably
destroy central part of the density profile. An immediate
example comes to mind is the Cassiopeia A supernova remnant
which appears more or less spherical with a nonrotating
X-ray bright neutron star left behind (i.e., no X-ray
pulsations are detectable so far by {\it Chandra} satellite
observations).

For `quasi-static' asymptotic solution (\ref{complex1})
and (\ref{complex2}) of the second type (with logarithmic
oscillations) at small $x\equiv r/(k^{1/2}t^n)$, we then
have the radial flow speed $u=k^{(1-K_1)/2}t^{-n(K_1-1)-1}r^{K_1}
[L_1\cos(K_2\ln x)-L_2\sin(K_2\ln x)]$,
the mass density $\rho\propto r^{-2/n}+\eta t^{(5n-4)/2}r^{-5/2}
\{[L_1(1+K_1)+L_2K_2]\cos(K_2\ln x)
-[L_2(1+K_1)-L_1K_2]\sin(K_2\ln x)\}$, the enclosed mass
$M\propto r^{3-2/n}$, the core mass accretion rate
$\dot M\propto k^{(1-K_1)/2}t^{-n(K_1-1)-1}r^{1/2}
[L_1\cos(K_2\ln x)-L_2\sin(K_2\ln x)]$,
where $\eta$ is a readily identifiable coefficient, $L_1$
and $L_2$ are two small parameters, $K_1=2/n-3/2$,
$K_2=(-1/4+6/n-4/n^2)^{1/2}$ and $n>2/3$. In reference to the
first type of self-similar `quasi-static' asymptotic solutions
(\ref{static8}), the qualitative distinction of this second
type of self-similar `quasi-static' asymptotic solutions
(\ref{complex1}) and (\ref{complex2}) is of course the
oscillatory feature. The overall strength of this oscillatory
feature tends to die out with increasing time $t$. Nevertheless,
at a given time $t$, the mass density profile of $\rho$ may still
retain a considerable oscillatory feature especially for small
$r$ or small $x$. Again, we propose to use the data of the
Cassiopeia A supernova remnant or other suitable supernova
remnants to identify or search for such oscillatory features.
%The radial inflow speed approaches zero towards the
%core center while the mass density diverges there.
A series of our self-similar shock solutions with `quasi-static'
asymptotic solutions of first and second types are consistent,
at least qualitatively, to the major features of such a plausible
rebound shock scenario for supernova explosions.

To be more specific and tangible, we present a series of
self-similar rebound shock solutions approaching central
`quasi-static' asymptotic solutions in Figure \ref{PolySample}.
They serve as examples of illustration for a possible hydrodynamic
similarity model of a rebound shock in a core collapsing progenitor
star. Among various physical possibilities, these rebound shocks
are constructed as such, because we would like to demonstrate
similarity shock solutions with (i) an inflow right on the
upstream side of the shock, which accounts for the creation
of a rebound shock propagating outward through a collapsing
envelope; (ii) a self-similar shock, modelling a rebound
shock after encountering a highly compressed over-dense core in a
gravitational collapse; (iii) a massive outward expansion behind
(downstream of) the rebound shock, such that gas materials of the
massive progenitor star are driven out by such a powerful bouncing;
and (iv) a `quasi-static' asymptotic behaviour at small $x$, so
that eventually the dense inner core comes to a near equilibrium
as time goes on. These four features are qualitatively consistent
with the rebound shock scenario of a perceived supernova explosion,
leaving a compact object behind. We show shock solutions with both
type I (solutions 3 and 4) and type II (solutions 1 and 2)
`quasi-static' asymptotic solutions at small $x$, because both
satisfy the above intuitive physical requirements qualitatively.
Solutions 1 and 3 have inflowing outer portions, while solutions
2 and 4 have outflowing outer portions\footnote{The progenitor
star may oscillate radially prior to and during the onset of a
gravitational core collapse as already noted earlier.} to
encompass various conceivable situations. In principle, there may
also exist contracting and breezing outer portions with other
features much alike the core collapse situations just described above.

\subsection[]{Spatial and Temporal Cutoffs}

For a massive progenitor star, the self-similar evolution of
a core collapsing process with a rebound shock cannot persist
all the time and thus a temporal or spatial cutoff needs to be
specified. However, the similarity
model is characterized by a continuous
mass density profile with no natural cutoffs present in solutions
themselves. We thus refer to astronomical data to estimate a
sensible boundary for our model application. In this model, we set
a radius $r_i$ (for instance, $r_i=10^6$cm if the compact object
is a neutron star or black hole, and $r_i=10^8$cm if it is a
white dwarf) for our sample solutions; and we choose a radius
$r_o$ (for example, $r_o\equiv10^{12}$cm) as the outer boundary
for application of our sample solutions. A typical massive star
has a radius of the order of $\simeq10^{12}$cm (see Herrero et
al. 1992 and Sch\"onberner \& Harmanec 1995 for observational
data on masses and radii of main-sequence O and B stars). In
principle, we could invoke even larger stellar radii for even
more massive stars.

For these spatial cutoffs, the cutoff time is introduced below
accordingly. A rebound shock emerges roughly around the core
of a massive star when the collapse produces an over-dense
degenerate core. It travels outward and evolves into a
self-similar phase. We take the time when the shock crosses
the inner radius (say, $\sim 10^6$cm for a neutron star radius)
as the initial time to apply our shock solution. This time
would then correspond to
\begin{equation}\label{model2}
t_1=\bigg(\frac{r_i}{k_1^{1/2}x_{s_1}}\bigg)^{1/n}\ .
\end{equation}
%waiting for confirmation from weigang Wang July 16, 2006
Here, $t_1$ is obtained assuming our model to be valid
from the beginning at $t=0$. This estimate also roughly
represents the time needed for the shock to evolve from
the center to the inner reference radius $r_i$. Specific
estimates of $t_1$ are summarized in Table \ref{tMasses}
with adopted parameters indicated. Since our `quasi-static'
asymptotic solution approaches a static state as
$t\rightarrow +\infty$, we do not set an upper
bound in time for our rebound shock model.

\begin{figure}
\includegraphics[scale=0.45]{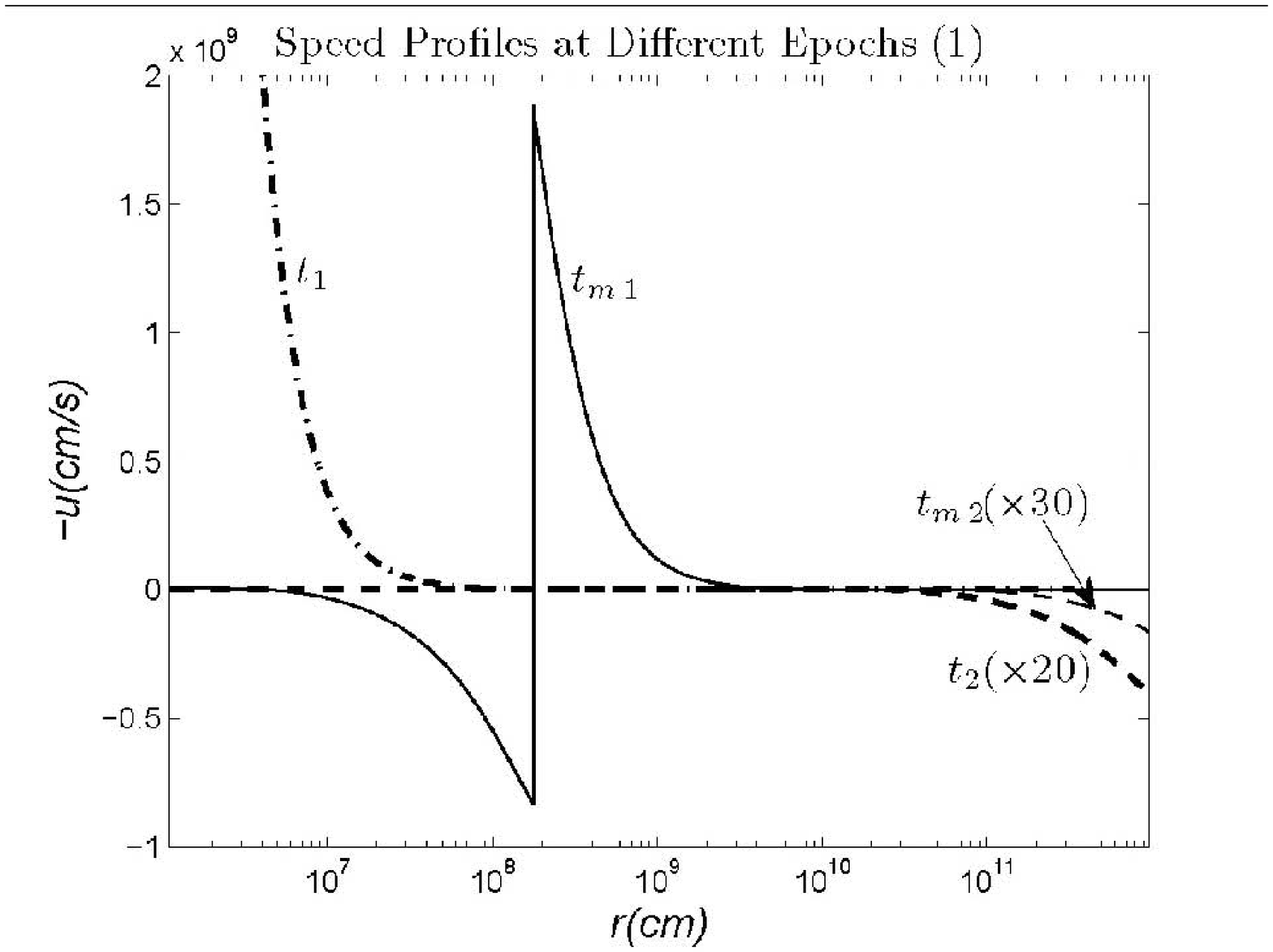}
\caption{The negative radial flow speed profile $-u$ at different
time $t$ values of the numerical example, with $r$ in the
logarithmic scale, for the model 1 in Figure \ref{PolySample}.
Here, $t_1=6.15\times10^{-5}$s is the time when the shock crosses
the inner radius $r_i$, $t_2=2.29\times10^4$s is the time when the
shock crosses the outer boundary $r_o$;
%$t_3=1.79\times10^5$s is the time when the vanishing
%point $x_{v=0}=0.03842$ crosses the outer boundary
$t_{m1}=0.1$s and $t_{m2}=10^5$s are intermediate times between
$t_1$ and $t_2$, and between $t_2$ and $t=\infty$, respectively.
Negative radial speed profiles have been multiplied by various
numerical factors shown in the figure for the compactness and
clarity of presentation.
%{\bf Please change the title to Speed Profiles at Different Epochs.}
} \label{PolyVelocity1}
\end{figure}

\subsection[]{Properties of Rebound Shock Solutions}

We note a common property of self-similar rebound shock solutions
shown in Figure \ref{PolySample}. As $x\rightarrow +\infty$ and
as $x\rightarrow0^{+}$, the profiles of radial flow speed $u$ and
mass density $\rho$ both become independent of time $t$. For a
specified radial range of $r$ within a star, the stellar
configurations are either static or stationary both at the
beginning ($t\rightarrow0^+$) and at the end
($t\rightarrow +\infty$). In particular, the enclosed
mass is given by
\begin{equation}\label{property1}
M=\frac{nk^{1/n}A}{(3n-2)G}r^{3-2/n}\
\end{equation}
for asymptotic solutions at large $x$, whereas
\begin{equation}\label{property2}
M=\frac{n}{(3n-2)G}\bigg[\frac{k2\gamma(3n-2)}
{n^{2}}\bigg]^{1/n}r^{3-2/n}\
\end{equation}
for the `quasi-static' asymptotic
solutions of both types at small $x$.

These analytical results may be utilized to grossly estimate
mass variations of a star during the core collapse and
supernova explosion processes. As $t$ approaches $0^{+}$, if
one takes our model to be valid from the beginning, then the
enclosed stellar mass $M$ is known once a progenitor stellar
radius $r_o$ is specified. Likewise, if one takes our model
to be valid after a sufficiently long time (i.e.,
$t\rightarrow +\infty$), a quasi-static core equilibrium is
ultimately achieved and the enclosed mass $M$ is known for
a specified core radius $r_i$.

A summary of the estimated enclosed mass values for both
the outer stellar radius $r_o$ and the inner reference
radius $r_i$ for solutions presented above are given in
Table \ref{kMasses} for the four rebound shock solutions
in Fig. \ref{PolySample}. We refer to the enclosed mass
at $t\rightarrow 0^+$ as the initial mass (hence the subscript
$_{ini}$), and the enclosed mass at $t\rightarrow +\infty$ as
the ultimate mass (hence the subscript $_{ult}$), respectively.
The initial mass may roughly represent the mass of a progenitor
star just evolved into a self-similar phase after the
initiation of a core collapse, and the ultimate mass may roughly
correspond to the mass of the remnant compact object as a long
time has elapsed. The determination of the $k$ values in both
the upstream (subscript $1$) and downstream (subscript $2$)
regions are described in the next subsection; while the inner
and outer radii are set to be $r_i=10^6$cm and $r_o=10^{12}$cm
here for considering remnant neutron stars or black holes.

\begin{table}
\center\caption{The adopted $k$ values and the initial and
ultimate enclosed mass values for the four models shown in
Figure \ref{PolySample}. The subscript $o$ denotes the outer
enclosed mass, and subscript $i$ denotes the inner enclosed
mass; while subscript $_{ini}$ represents the initial mass,
and $_{ult}$ represents the ultimate mass.}\label{kMasses}
\begin{tabular}{ccccc}\hline
No.&1&2&3&4\\
\hline
$\!\!\!\!\!$$k_1$(cgs units)$\!\!\!$&$\!\!$$8.87\times10^{14}$$
\!\!$&$\!\!\!\!\!$$3.21\times10^{16}$$\!\!\!$&$
\!\!\!3.36\times10^{16}\!\!\!$&$\!\!1.735\times10^{17}\!\!\!$\\
$\!\!\!\!\!$$k_2$(cgs units)$\!\!\!$&$\!\!$$3\times10^{17}$$
\!\!$&$\!\!\!\!\!$$3\times10^{17}$$\!\!\!$&$
\!\!\!3\times10^{17}\!\!\!$&$\!\!3\times10^{17}\!\!\!$\\
$\!\!\!\!\!$$M_{i,ini}(M_{\odot})$ $\!\!\!$&$\!\!$3.35$\!\!$&$
\!\!\!\!\!$3.38$\!\!\!$&$\!\!\!$3.11$\!\!\!$&$\!\!$3.18$\!\!\!$\\
$\!\!\!\!\!$$M_{i,ult}(M_{\odot})$ $\!\!\!$&$\!\!$1.42$\!\!$&$
\!\!\!\!\!$1.42$\!\!\!$&$\!\!\!$1.65$\!\!\!$&$\!\!$1.65$\!\!\!$\\
$\!\!\!\!\!$$M_{o,ini}(M_{\odot})$ $\!\!\!$&$\!\!$24.1$\!\!$&$
\!\!\!\!\!$24.3$\!\!\!$&$\!\!\!$7.02$\!\!\!$&$\!\!$7.17$\!\!\!$\\
$\!\!\!\!\!$$M_{o,ult}(M_{\odot})$ $\!\!\!$&$\!\!$10.2$\!\!$&$
\!\!\!\!\!$10.2$\!\!\!$&$\!\!\!$3.71$\!\!\!$&$\!\!$3.71$\!\!\!$\\
\hline
\end{tabular}
\end{table}

\subsection[]{Estimates for values of $k$ Parameter}

Clearly, once the numerical value of the $k$ parameter
in similarity transformation (\ref{transform1}) and
(\ref{transform2}) is determined for a star, the shock
model is then specified. Since
\[
k=\frac{p}{\rho^{\gamma}(4\pi G)^{\gamma-1}}
\]
\begin{equation}\label{model1}
\qquad=\frac{Nk_BT}{\rho^{\gamma}(4\pi G)^{\gamma-1}}
=\frac{k_BT}{\bar{m}\rho^{\gamma-1}(4\pi G)^{\gamma-1}}\ ,
\end{equation}
where $N$ is the particle number density, $\bar{m}$ is the mean
molecular (atomic) mass of the gas particle, and the second
equality holds only for an ideal gas, we may in principle
determine $k$ parameter using the physical quantities in equation
(\ref{model1}). Nevertheless, even for long-studied main-sequence
stars, the determination of physical parameters in the stellar
interior has been challenging, not to mention the relevant
parameters for massive collapsing stars, which involve various
rapid physical processes and are extremely rare to be caught in
action. Currently, some theoretical estimates of such parameters
are available, although most of these parameters are estimated
for static configurations, e.g., main-sequence stars or neutron
stars. These two stellar configurations as part of our model
considerations, are the progenitor star and the remnant compact
object left behind a rebound shock, respectively. While the
value of $k$ parameter is by no means constant during complicated
processes before and after the similarity evolution involving a
constant $k$, we would presume for simplicity that the $k$ value
does not change significantly during the dynamical evolution.

%{\it
\cite{AppenzellerTscharnuter74}, \cite{bond84},
\cite{NadyozhinRzinkova2005} and \cite{schaller92} gave
theoretical estimates for the hydrogen-burning phase of massive
stars with $\rho_c\sim$ 1g/cm$^3$ and $T_c\sim10^{7-8}$K.
Using $\bar m=0.5m_p$ with $m_p$ being the proton mass, we
obtain approximately $k\sim 10^{17}$ cgs units. \cite{Arnett77},
%\cite{bond84}
Bond et al. (1984), and \cite{schaller92} provided theoretical
predictions for the late evolution phase of massive stars
after the hydrogen-burning phase, and the results are
$\rho_c\sim 10^8$g/cm$^3$ and $T_c\sim 10^9$K. Using these
late-phase parameters and $\bar m=4m_p/3$ for helium atoms,
we derive $k\sim10^{16}$ cgs units. Also from the equation
of state for degenerate neutrons, the corresponding
$p_c\sim10^{35}$dynes/cm$^2$ and $\rho_c\sim10^{15}$g/cm$^3$
may be appropriate for neutron stars (e.g., Shapiro \&
Teukolsky 1983), and the estimated $k$ value is accordingly
$\sim10^{17}$ cgs units.

In the rebound shock model construction as we intend to
discuss applications to SNe and the formation of remnant
neutron stars, we may first assign $k_2$ values for a
reasonable enclosed mass range around $\sim 1M_{\odot}$.
To be consistent with estimates for typical neutron stars,
the $k_2$ value is estimated to be $\sim 3\times10^{17}$cgs
units for both type I and type II `quasi-static' asymptotic
solutions at small $x$. The value of $k_1$ is derived
accordingly for each shock model, using the $\lambda$
parameter from the shock jump conditions. The $k$ values
thus determined in our shock models are summarized in
Table \ref{kMasses}. We see that all the $k_1$ values are
of the orders of $\sim 10^{15\sim 17}$cgs units, fairly
close to the estimates for massive progenitor stars.

\begin{figure}
\includegraphics[scale=0.45]{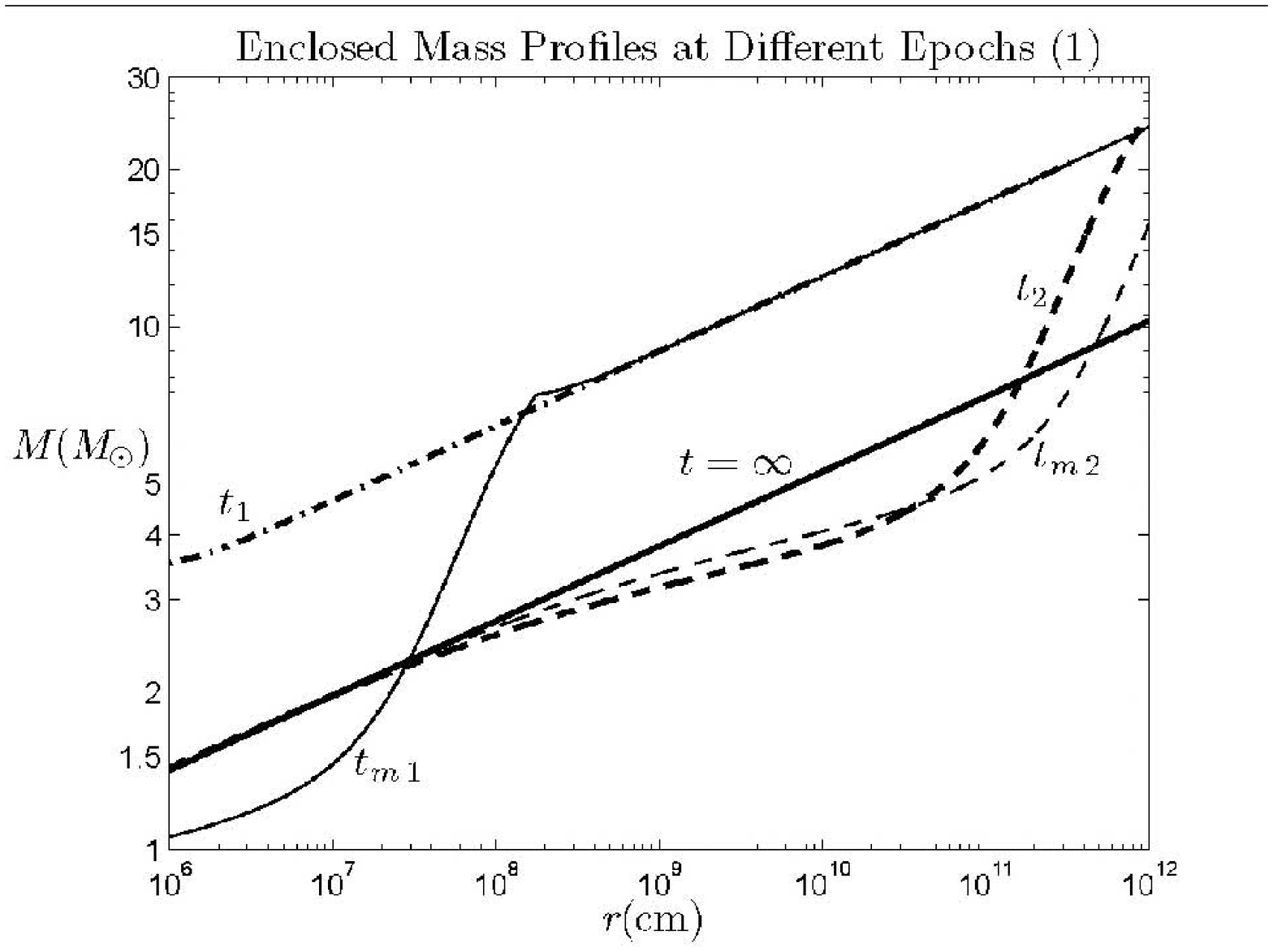}
\caption{The total enclosed mass at different time $t$ values
of the numerical example in logarithmic scales, for the first
model in Figure \ref{PolySample}. Here, $t_1=6.15\times10^{-5}$s
is the time when the shock crosses the inner radius $r_i$,
$t_2=2.29\times10^4$s is the time as the shock crosses the
outer boundary $r_o$;
%$t_3=1.79\times10^5$s is the time when the vanishing
%point $x_{v=0}=0.03842$ crosses the outer boundary
$t_{m1}=0.1$s and $t_{m2}=10^5$s are intermediate time
values between $t_1$ and $t_2$, and between $t_2$ and
$t=\infty$, respectively.
%{\bf Please change the title to Enclosed
%Mass Profiles at Different Epochs. }
}\label{PolyMass1}
\end{figure}

\begin{figure}
\includegraphics[scale=0.45]{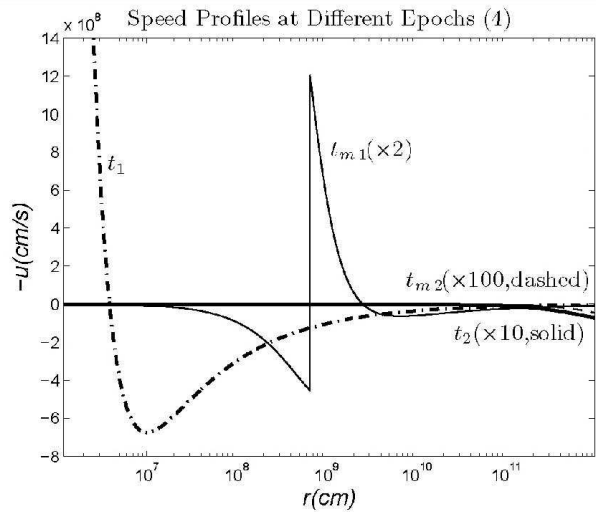}
\caption{The radial speed profile at different time $t$ values
of the numerical example in logarithmic scales for the fourth
model in Figure \ref{PolySample}. Here, $t_1=7.18\times10^{-5}$s
is the time as the shock crosses the inner radius $r_i$,
$t_2=4.78\times10^4$s is the time as the shock crosses the outer
boundary $r_o$;
%$t_3=1.79\times10^5$s is the time when the vanishing
%point $x_{v=0}=0.03842$ crosses the outer boundary
$t_{m1}=1$s and $t_{m2}=10^6$s are intermediate times between
$t_1$ and $t_2$, and between $t_2$ and $t=\infty$, respectively.
The speed profiles have been multiplied by various factors as
shown for the compactness and clarity of the presentation,
%{\bf Please change the title to Speed Profiles at Different Epochs.}
}\label{PolyVelocity4}
\end{figure}

\begin{table}
\center\caption{Below all time $t$ in seconds and all enclosed
mass in unit of the solar mass $M_{\odot}$. Here, $t_1$ for the
time as the shock reaches the inner boundary and $t_2$ for the
time as the shock reaches the outer boundary and the corresponding
$M_i$ and $M_o$ for the four models in Figure \ref{PolySample}.
Enclosed masses for $t_{m1}$ and $t_{m2}$ as indicated in relevant
figures are also included. The $k$ parameters are according to
Table \ref{kMasses}.}\label{tMasses}
\begin{tabular}{ccccc}\hline
No.&1&2&3&4\\
\hline
$t_1$&6.15$\times10^{-5}$&6.04$\times10^{-5}$
&8.16$\times10^{-5}$&7.18$\times10^{-5}$\\
$t_2$&2.29$\times10^4 $ &2.25 $\times10^4$
&5.44$\times10^{4} $&4.78$\times10^{4}$\\
$M_{i1} $&3.53&3.54&3.22&3.23\\
$M_{i2} $&1.43&1.43&1.69&1.69\\
$M_{o1} $&24.1&24.3&7.03&7.17\\
$M_{o2} $&25.4&25.4&7.24&7.28\\
$M_{im1}$&1.06&1.06&1.85&1.85\\
$M_{im2}$&1.43&1.43&1.68&1.68\\
$M_{om1}$&24.1&24.3&7.03&7.17\\
$M_{om2}$&15.6&15.6&5.78&5.78\\
\hline
\end{tabular}
\end{table}

\subsection[]{Shock Model Presentation}

\begin{figure}
\includegraphics[scale=0.45]{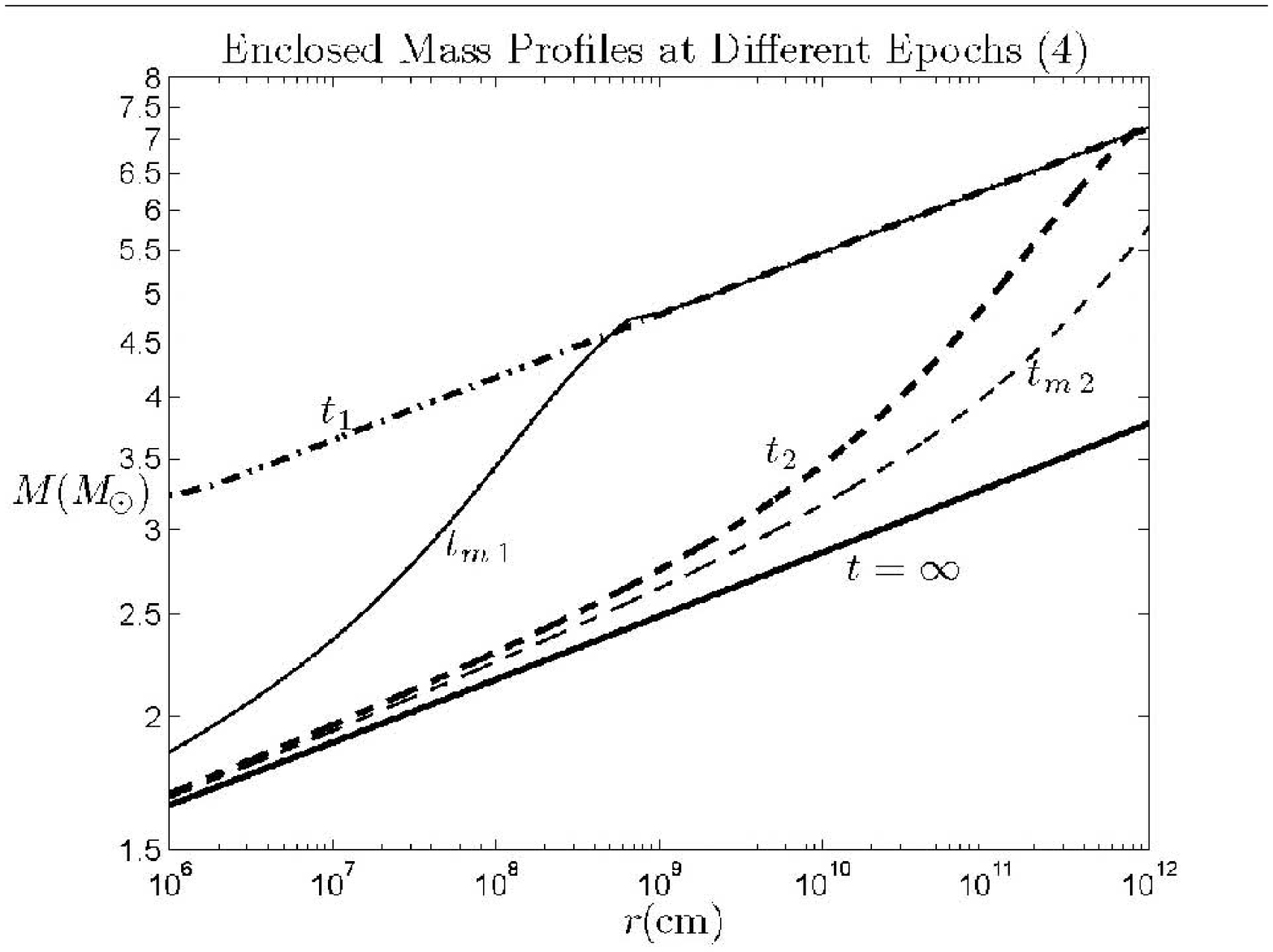}
\caption{The total enclosed mass at different time $t$ values of
the numerical example, in logarithmic scales for the first model
in Figure \ref{PolySample}. Here, $t_1=7.18\times10^{-5}$s is
the time when the shock crosses the inner reference radius $r_i$,
$t_2=4.78\times10^4$s is the time as the shock crosses the outer
boundary $r_o$;
%, $t_3=1.79\times10^5$s is the time when the vanishing
%point $x_{v=0}=0.03842$ crosses the outer boundary
$t_{m1}=11$s and $t_{m2}=10^6$s are intermediate
time values between $t_1$ and $t_2$, and between
$t_2$ and $t=\infty$, respectively.
%{\bf Please change the title to Enclosed
%Mass Profiles at Different Epochs. }
}\label{PolyMass4}
\end{figure}

Figs. \ref{PolyVelocity1} and \ref{PolyMass1}, as well
as Figs. \ref{PolyVelocity4} and \ref{PolyMass4} present
the profiles of the negative radial flow speed $-u$ and
the enclosed mass $M$ at different temporal epochs,
respectively.

Besides the timescale $t_1$ as the initial time of our
model, we introduce another typical timescale $t_2$
which is the time when the shock reaches the outer
radius $r_o$, namely
\begin{equation}\label{model3}
t_2=\bigg(\frac{r_o}{k_1^{1/2}x_{s_1}}\bigg)^{1/n}\ .
\end{equation}
We presume our model to be valid since $t\rightarrow 0^+$.
The specific $t_2$ values relevant to various shock models
are also contained in Table \ref{tMasses} with adopted
parameters clearly indicated. As a rebound shock cannot be
seen from the Earth until the shock reaches the stellar
photosphere and our outer boundary is set for the radius
of a massive star, this time value roughly corresponds to
the time for a rebound shock to travel out and be seen.
%At this time, the shock has driven most materials out of the
%stellar interior, but the accretion has continued at the
%stellar boundary. Thus the core mass has decreased and the
%total enclosed mass within $r_o$ has increased. The heavy
%dashed curves in Figures \ref{PolyVelocity1}$-$\ref{PolyMass4}
%correspond to this time $t_2$.

Quantities at five different epochs: $t_1$, $t_2$, $t=\infty$
and two intermediate times $t_{m1}$ and $t_{m2}$ are shown in
the radial range $r_i=10^6\mbox{cm}<r<r_o=10^{12}$cm in Figs.
\ref{PolyVelocity1}, \ref{PolyMass1}, \ref{PolyVelocity4} and
\ref{PolyMass4}. The corresponding radial flow speeds of model
2 are like those of model 4, and the corresponding radial flow
speeds of model 3 are like those of model 1. Meanwhile, the
corresponding enclosed mass distributions of model 2 are like
those of model 1, and the corresponding enclosed mass
distributions of model 3 are like those of model 4.
%{\bf Sure of no disorder?} {\bf Confirmation from wwg.}

\subsection[]{Analysis of the Rebound Shock Model}

We now describe model 1 results as an example of illustration.
Relevant data of this model as well as other models are summarized
in Tables \ref{kMasses} and \ref{tMasses}. Initially, the stellar
interior is in a core collapse (inflow) stage, which is triggered
by the exhaustion of the central nuclear fuel. The initial profile
is also significantly over-dense as compared to the eventual
quasi-static configuration after a long lapse in time. The outer
envelope remains collapsing as the rebound shock emerges around
the central degenerate core and travels outward. The shock travels
faster than the sound speed and the perturbation information cannot
reach the outer region until the shock reaches there. At time $t_1$,
the enclosed mass within $r_o$ is $24.1M_{\odot}$ and within $r_i$,
it is $3.53M_{\odot}$.

The rebound shock then breaks out from the stellar interior,
travelling at a fantastic speed (mostly in the range of $10^7
\sim 10^9$cm/s). During the first few seconds, the core mass
decreases rapidly. At time $t_{m1}=0.1$s, the enclosed mass
within $r_i$ has reduced to $1.06M_{\odot}$, while within $r_o$ the
enclosed mass is still $24.1M_{\odot}$ because the time difference
is small and the mass accretion is not significant. When the shock
reaches the outer boundary, a slight accretion has changed the total
enclosed mass within $r_o$ to $25.4M_{\odot}$, yet the inner core
has already experienced a minor accretion due to radial speed
oscillation of our type II `quasi-static' asymptotic solution, and
the enclosed mass within the inner radius $r_i$ has now increased
to $1.43M_{\odot}$.

After the rebound shock passes over
the outer reference radius $r_o$,
%{\bf typo in subscript? $e$?}{\it sorry, it is.},
it is now the time for the massive star to drive out a significant
amount of mass. As the drive gradually weakens, the total enclosed
mass within $r_o$ at time $t_{m2}=10^5$s falls to $15.6M_{\odot}$,
while within $r_i$ the total enclosed mass remains $1.43M_{\odot}$
as before. Eventually the enclosed mass within $r_o$ becomes
$10.2M_{\odot}$, and within $r_i$, it becomes $1.42M_{\odot}$.
%Although after $t_3$, the total enclosed mass within $r_e$ may
%slightly increase because of the mass accretion if we further
%apply the self-similar solution to this model, we shall call
%it the end of our application because the mass driving-out is
%over, and the shock is now far away from the outer boundary.
From the core mass variation and the remaining
core mass within the inner reference radius $r_i$
in the end, we reasonably conjecture that the
remnant core may become a proto-neutron star.
%This idealized model possessing major desired
%properties of forming a neutron star, may be
%utilized as a self-similar model for this process.
The rebound shock velocity also conforms with usual
estimates.

\subsection[]{Mass Ratio Evolution}

One interesting aspect of our rebound shock model is the ratio
of the initial outer mass $M_{o,ini}$, roughly corresponding
to the initial mass of a prescribed progenitor star, and the
ultimate inner mass $M_{i,ult}$, roughly corresponding to the
mass of a remnant compact object in the core. By equations
(\ref{property1}) and (\ref{property2}), this mass ratio is
simply
\begin{equation}\label{massdrive1}
\frac{M_{o,ini}}{M_{i,ult}}=\frac{A}{\lambda^{2/n}}
\bigg[\frac{n^2}{2\gamma(3n-2)}\bigg]^{1/n}
%\cdot
\bigg(\frac{r_o}{r_i}\bigg)^{(3-2/n)}\ .
\end{equation}
In expression (\ref{massdrive1}), the last factor involving the
ratio $r_o/r_i$ which is somewhat arbitrary, although we can
estimate proper radii for different progenitor stars and ultimate
compact core objects in a sensibel range. The ratio of the two
radii naturally affects the mass ratio, but as $n$ approaches
$2/3$, or equivalently $\gamma$ approaches $4/3$, this last factor
would approach unity. The two factors together in front of this
somewhat arbitrary last factor in expression (\ref{massdrive1})
is the ratio of the initial mass and the final mass within the
same radius (i.e., the last factor is 1 in this case). This can
vary when $\gamma$ or equivalently $n$ varies, but it also depends
on $x_{s_2}$ or $(x_0\ ,v_0\ ,\alpha_0)$, which is the point on the
sonic critical curve from which we construct the `quasi-static'
portion of our rebound shock solutions (see subsection 6.2). Since
the relation of $\lambda$ and $A$ is explored numerically, we here
only provide the corresponding results for several cases with
$n=0.68$ (see Table \ref{drivingratio}), and the cases with other
$n$ or $\gamma$ values are fairly similar.

According to numerical results of Table \ref{drivingratio},
mass ratio (\ref{massdrive1}) increases for smaller $\alpha_0$
or for larger $x_{s_2}$. This is conceivable, in the sense that
as $\alpha_0$ becomes smaller, the `quasi-static' portion of the
rebound shock solution has a higher $v$ for more mass being
driven out; and when $x_{s_2}$ becomes smaller, the outer portion
has a higher $v$ for more mass being driven out. Referring to
Table \ref{drivingratio} and the relevant definitions therein, we
also infer empirically that ratio 1 may approach 1, but whether
this ratio 1 can be arbitrarily large is unclear. It appears that
the choice of self-similar rebound shock solutions affects the
mass ratio in a significant manner. We infer from this feature
that in reality the mass ratio may also vary, and thus a compact
object with a definite mass may possibly come from progenitor
stars with different initial masses.

\begin{table}
\center\caption{Evolution of mass ratio $M_{o,ini}/M_{i,ult}$ for
various parameters when $\gamma=1.32$ and $n=0.68$ are specified.
For each value of $\alpha_0$, the point from which we construct
our rebound shocks, we explore two cases, namely, rebound shock
solutions with contracting and expanding envelopes: in the
neighbouring two sets with the same $\alpha_0$ values, the one
with smaller $x_{s_2}$ gives an expanding envelope, and the one
with smaller $x_{s_2}$ gives a contracting envelope. In this
Table, ratio 1 is the first two factors together in expression
(\ref{massdrive1}), namely, $A\big\{n^2/[\lambda^22\gamma(3n-2)]
\big\}^{1/n}$, and ratio 2 is the last factor in expression
(\ref{massdrive1}), namely, $\big(r_o/r_i\big)^{(3-2/n)}$.}
\label{drivingratio}
\begin{tabular}{cccccc}\hline
$\alpha_0$&$x_{s_2}$&$\lambda$&$A$&ratio 1&ratio 2\\
\hline
0.067&1.2&1.315&0.493&1.933&4.357\\
0.067&1.1&2.989&5.401&1.893&4.266\\
0.070&1.1&1.064&0.2048&1.497&3.373\\
0.070&0.9&1.737&0.8338&1.442&3.250\\
0.072&1&1.0628&0.1808&1.326&2.989\\
0.072&0.8&1.843&0.879&1.277&2.878\\
0.074&1&1.0248&0.149&1.216&2.742\\
0.074&0.8&1.262&0.2662&1.178&2.655\\
0.076&1&1.0100&0.1326&1.130&2.546\\
0.076&0.8&1.1238&0.1766&1.099&2.478\\
0.078&1&1.00383&0.1219&1.0575&2.384\\
0.078&0.8&1.0663&0.1422&1.0329&2.328\\
0.079&1&1.0000235&0.1175&1.0308&2.323\\
0.079&0.8&1.0496&0.1320&1.00440&2.264\\
\hline
\end{tabular}
\end{table}

\section[]{Remnant Proto-White Dwarfs}

Conventionally, one does not invoke a rebound shock
explosion to expose the degenerate core, a proto-white
dwarf, of a progenitor star. It is a general belief that
various mass loss mechanisms through winds are responsible
for eventually producing or exposing proto-white dwarfs.
However, given our theoretical framework outline in this
paper, it is not obvious why a rebound shock cannot occur to
blow away stellar envelope and produce a proto-white dwarf.
After all, a progenitor star of mass $\sim 6-8M_{\odot}$
has to throw away a mass amount of $\sim 4.6-6.6M_{\odot}$
in order to expose a proto-white dwarf in the core.

To consider the formation of a proto-white dwarf in the
degenerate core of a progenitor star in the high-mass end
(say, $\sim 6-8M_{\odot}$), we would set $r_i$ to be a typical
radius of several thousand kilometers (say, $\sim 3\times 10^8$cm)
for a white dwarf and $r_o\sim 10^{13}$cm to be the stellar radius
to enclose a mass in the range of $\sim 6-8M_{\odot}$ for the
corresponding progenitor star. After the initiation of core collapse
and as a powerful self-similar rebound shock travels outward, the
masses enclosed within $r_i$ and $r_o$ decrease, sometimes oscillate,
and eventually approach finite values with the increase of time $t$.
If the final mass $M_{i,ult}$ within $r_i$ is less than the
Chandrasekhar mass limit of $1.4M_{\odot}$, we would say that a
proto-white dwarf is produced in the remnant core. Otherwise if the
final mass $M_{i,ult}$ within $r_i$ is greater than $1.4M_{\odot}$,
say $\sim 2-3M_{\odot}$, we would then hypothesize that dynamical
instabilities associated with a proto-white dwarf exceeding the
Chandrasekhar mass should have already happened before reaching
the final model state of $t\rightarrow +\infty$ or $x\rightarrow 0^+$;
for this to occur, the timescales of such dynamical instabilities
should be shorter than the timescale of self-similar evolution
around $r_i$. Such dynamical instabilities can give rise to a
neutron star. This might then be viewed as a two-stage process
to produce a neutron star with a progenitor mass roughly in the
range of $\sim 6-8M_{\odot}$ or higher.

For progenitor stars with masses less than $\sim 6-8M_{\odot}$,
similar rebound shocks may occur but with weaker strengths. Such
rebound shocks could fail to do the driving completely as a result
of various energy losses, yet their presence and signatures may
be detectable.

We now explore a few more examples for producing proto-white
dwarfs or proto-neutron stars in core collapse and rebound
shock processes. For all these cases below, we now take
$r_i=3\times 10^8$ cm and $r_o=10^{13}$ cm.

The relevant parameters for the first rebound shock model are:
$n=0.7$ (or equivalently $\gamma=1.3$), $\alpha_0=0.175$,
$x_0=1.672$, $v_0=0.293$, $x_{s_2}=1.1$, $\lambda=1.133$,
$A=0.752$, $B=-0.222$, $k_2=1.65\times 10^{17}$ cgs unit,
$k_1=1.29\times 10^{17}$ cgs unit, $M_{o,ini}=7.85M_{\odot}$,
$M_{i,ult}=1.36M_{\odot}$. This is a case of producing a
remnant proto-white dwarf near the Chandrasekhar mass limit
of $M_{i,ult}=1.36M_{\odot}$ from a progenitor star of mass
$M_{o,ini}=7.85M_{\odot}$. We feel that with more physical
input and requirements, our model can provide sensible dynamic
constraints on the initial and final masses of progenitor and
of remnant proto-white dwarf.

The relevant parameters for the second rebound shock model are:
$k_2=2.2\times 10^{17}$ cgs unit, $k_1=1.71\times 10^{17}$ cgs
unit, $M_{o,ini}=11.8M_{\odot}$, $M_{i,ult}=2.05M_{\odot}$;
other parameters are the same as the first model parameters
above. As already noted earlier, this proto-white dwarf
may be subject to dynamical instabilities and could lead
to the formation of a proto-neutron star in the end.

The relevant parameters for the third model are: $n=0.7$ (or
equivalently $\gamma=1.3$), $\alpha_0=0.196$, $x_0=1.324$,
$v_0=0.0341$, $x_{s_2}=0.8$, $\lambda=1.146$, $A=0.5585$,
$B=-0.864$, $k_2=2.1\times 10^{17}$ cgs unit, $k_1=1.60
\times 10^{17}$ cgs unit, $M_{o,ini}=7.96M_{\odot}$,
$M_{i,ult}=1.92M_{\odot}$. Again, this proto-white dwarf
may be subject to dynamical instabilities and could lead
to the formation of a proto-neutron star eventually.

\section[]{Discussion and Speculations}

We are able to construct various semi-complete self-similar
solutions either with or without a rebound shock; these
solutions may be adaptive to various astrophysical settings.
Our numerical examples may be interpreted as an approximate
model of a rebound shock initiated during the gravitational
core collapse in a massive star, which drive materials out
to eventually form possible quasi-static configurations.

For the convenience and clarity of discussion, we mainly
focus on rebound shocks for supernovae and remnant neutron
stars in the preceding sections.
In fact, we may well choose a larger value of $k_2$
for shock models 3 and 4 in Figure \ref{PolySample}, say,
5 times larger, it follows that all the masses and radial
flow speeds should be multiplied by the same factor of 5.
Then the initial progenitor mass for these two shock models
would be around $35M_{\odot}$ and the final mass enclosed
within $r_i$ would be $8.25M_{\odot}$. This then implies a
black hole left behind as the remnant compact object during
the core collapse and rebound shock processes\footnote{Brown
\& Bethe 1994 suggested supernova produced low-mass black
holes with masses only slightly above $1.5M_{\odot}$.}. In
addition, we can also choose relevant parameters to model
the formation of white dwarfs, for instance, taking
$r_i=3\times 10^8$ cm and $r_o=10^{13}$ cm, we can
construct an example for $M_{o,ini}=6.3M_{\odot}$ and
$M_{i,ult}=1.1M_{\odot}$. Although we provide quantitative
examples using our rebound shock model framework, our main
motivation is to point out various conceptual possibilities
seemingly physically plausible. After all, our current
hydrodynamic model is very limited in many ways.

While we cannot draw sure conclusions on the fate of
progenitor stars, we may outline a scheme for possible remnant
compact objects left behind based on the current theoretical
knowledge. Within the inner reference radius $r_i\sim 10^6$cm, the
final static configuration is assumed to be a proto-neutron star if
the enclosed mass is in the range of $\lsim 3-4M_{\odot}$ or a black
hole if the enclosed mass is higher than $\sim 4M_{\odot}$; by the
same token, within the inner reference radius $r_i\sim 3-4\times
10^8$cm, the final static configuration is assumed to be a
proto-white dwarf if the enclosed mass is in the range of $\lsim
1.4M_{\odot}$ or a proto-neutron star if the enclosed mass is
higher $\sim 1.4M_{\odot}$.

Our analysis on the sonic critical curve shows that the critical
curve has qualitative differences as compared to that of the
isothermal case (e.g., Shen \& Lou 2004), i.e., at small $x$ values
the sonic critical curve diverges here, and there exists a lower
limit of the reduced density $\alpha$ on the sonic critical curve.
Such analysis is technically useful in determining the sonic
critical curve, and is also adaptive in determining the
asymptotic behaviours of other complicated functions.

The `quasi-static' asymptotic solution is so named after the
first-order of this solution (viz., the semi-complete global SPS
solution \ref{pre1} or \ref{static1}). This new asymptotic solution
at small $x$ is a characteristic of polytropic flows only. For the
first type of this `quasi-static' asymptotic solution with a real
$K>1$, since $L$ can be either positive or negative, the
perturbation on the SPS can be either outflow or inflow and
influences the density
profile correspondingly (density perturbation is positive for an
outflow and negative for an inflow). The second type of this
`quasi-static' asymptotic solution with complex $K$ represents
a wave-like perturbation imposed on a SPS. Both the radial flow
speed and density profiles have an oscillatory feature, and
according to equations (\ref{complex1}) and (\ref{complex2}),
the oscillatory terms have a phase difference not larger than
$\pi/4$ between density and velocity profiles.
%(this phase difference is not larger than $\pi/4$).
The $x^{K_1}\exp(iK_2\ln x)$ term in radial flow speed profile
has an interesting property that the vibration decreases when
$x$ gets smaller.

In the context of supernova explosions, we propose to search
for characteristic signatures in density profiles of both
types of such `quasi-static' asymptotic solutions in the
Cassiopeia A supernova remnant as well as other suitable
supernova remnants. While unconventional, we suspect that
some proto-white dwarfs are also formed involving rebound
shocks and explosions.

Since various other physical conditions, such as the neutrino
opacity, radiation pressure, general relativistic effects,
magnetic field, and rotational effects are generally involved
in such core collapsing phase of stellar evolution, we hope
that this scenario and interpretation, while limited by a
highly idealized model framework, catch some essential
features of supernova explosion. This scenario should
be tested by observations and numerical simulations.
%Since radiation is involved in any model ready
%for comparison between observation and theory,
We plan to incorporate a random magnetic field into this model
framework. Thus a parallel analysis of a magnetohydrodynamic
model similar to \cite{YuLou05} and Yu et al. (2006) is anticipated.
%Also, neutron stars usually rotates rapidly, thus the model
%is lacking any information on this property, and a more detailed
%model should describe this behaviour as well. These relevant
%research topics will be pursued in separate papers.

\section*{Acknowledgments}
This research has been supported in part by the ASCI Center for
Astrophysical Thermonuclear Flashes at the University of Chicago,
%under the Department of Energy contract B341495,
by the Special Funds for Major State Basic Science Research
Projects of China, by the Tsinghua Center for Astrophysics,
by the Collaborative Research Fund from the National Science
Foundation of China (NSFC) for Young Outstanding Overseas
Chinese Scholars (NSFC 10028306) at the National Astronomical
Observatories, Chinese Academy of Sciences, by the NSFC grants
10373009 and 10533020 at the Tsinghua University, and by the
SRFDP 20050003088
%Specialized Research Fund for the Doctoral Program of Higher Education
and the Yangtze Endowment from the Ministry
of Education at the Tsinghua University.
%The hospitalities of the Mullard Space Science Laboratory at
%University College London, U.K., of School of Physics and
%Astronomy, University of St Andrews, Scotland, U.K., and of
%Centre de Physique des Particules de Marseille (CPPM/IN2P3/CNRS)
%et Universit\'e de la M\'editerran\'ee Aix-Marseille II, France
%are also gratefully acknowledged.
Affiliated institutions of Y-QL share this contribution.

\vskip 0.4cm

\appendix
\section[]{Polytropic Cases of $\gamma\geq4/3$}\label{gamma}

When index $\gamma>4/3$ is adopted for a polytropic gas, the
self-similar transformation should be modified slightly. In
fact, the only difference between this case and what we have
considered in the main text concerns the transformation for
the total enclosed mass $M$. As $M>0$ is a physical requirement
in transformation (\ref{transform2}), the scaling factor $e$
should take the form of
\begin{equation}\label{gamma1}
e=\frac{k^{3/2}t^{3n-2}}{(2-3n)G}\
\end{equation}
for $\gamma>4/3$ in a parallel analysis.
%and a corresponding analysis will be slightly
%different, but quite parallel with our case.
For example, the $nx-v>0$ criterion for $m>0$
should now be replaced by $nx-v<0$.

The $\gamma=4/3$ case is special and makes the scaling factor $e$
in definition (\ref{transform2}) meaningless. Noting that the
factor $3n-2$ was only added for convenience, we may simply set
\begin{equation}\label{gamma2}
e=\frac{k^{3/2}}{G}
\end{equation}
to perform a self-similar transformation. Substituting this
modified transformation into hydrodynamic equations (\ref{mass1})
through (\ref{state}) under spherical symmetry, we obtain
\begin{equation}\label{gamma3}
v=nx=2x/3\ ,
\end{equation}
\begin{equation}\label{gamma4}
m'=\alpha x^2\ ,
\end{equation}
and a second-order nonlinear ODE for $\alpha$ alone
\begin{equation}\label{gamma5}
\alpha''+\frac{2\alpha'}{x}-\frac{2(\alpha')^2}{3\alpha}
-\frac{1}{2}\alpha^{2/3}+\frac{3}{4}\alpha^{5/3}=0\ .
\end{equation}
%The solution properties of this ODE for $\alpha$ remain to be
%examined. While this case was mentioned by \cite{yahil83} as
%the homologous collapse of \cite{goldreich80}, it appears that
%the homologous collpase of \cite{goldreich80} does not have
%the property of $u\propto r/t$ valid in this $n=2/3$ case.
We shall pursue a more complete analysis of this problem
in a separate paper.

%Although a treatment on these cases shall not be complicated for
%us right now, because of lack of space, and for convenience of
%readers to focus on the main points of our current paper, we did
%not include a detailed analysis upon such cases.

\label{lastpage}


\vskip 0.4cm

\begin{thebibliography}{99}
\bibitem[\protect\citeauthoryear{Appenzeller \&
Tscharnuter}{1974}]{AppenzellerTscharnuter74}Appenzeller I.,
Tscharnuter W., 1974, A\&A, 30, 423
%-430, The Evolution of a Massive Protostar
\bibitem[\protect\citeauthoryear{Arnett}{1977}]{Arnett77}
Arnett W. D., 1977, ApJ Suppl. Ser., 35, 145
%-159, Advanced Evolution of Massive Stars. VII. Silicon Burning
\bibitem[\protect\citeauthoryear{Bethe}{1993}]{Bethe93}
Bethe H. A., 1993, ApJ, 412, 192
%--Hans, 202; SN 1987A: An Empirical and Analytic Approach
\bibitem[\protect\citeauthoryear{Bethe}{1995}]{Bethe95}
Bethe H. A., 1995, ApJ, 449, 714
%--726; The Supernova Shock
\bibitem[\protect\citeauthoryear{Bian \& Lou}{2005}]
{BianLou05}Bian F.-Y., Lou Y.-Q., 2005, MNRAS, 363, 1315
%--1328; Spherical Isothermal Self-Similar Shock Flows
%in directory BianFuYan/Note/draft.tex
%submitted on July 1, 2005 Friday THCA 166.111.26.56
\bibitem[\protect\citeauthoryear{Bionta et al.}{1987}]{Bionta87}
Bionta R. M., et al., 1987, Phys. Rev. Lett., 58, 1494
%IMB collaboration 8 neutrinos from SN 1987A.
\bibitem[\protect\citeauthoryear{Bodenheimer \&
Sweigart}{1968}]{bs68}Bodenheimer P., Sweigart A., 1968, ApJ, 152, 515
%Dynamic Collapse of the Isothermal Sphere
%
\bibitem[\protect\citeauthoryear{Boily \& Lynden-Bell}{1995}]{boily95}
Boily C. M., Lynden-Bell D., 1995, MNRAS, 276, 133
%self-similar collapse and accretion of radiative gas
%
%\bibitem[\protect\citeauthoryear{Bondi}{1952}]{bondi52}
%Bondi H., 1952, MNRAS, 112, 195
%%spherical accretion
%%
\bibitem[\protect\citeauthoryear{Bond, Arnett \& Carr}{1984}]{bond84}
Bond J. R., Arnett W. D., Carr B. J., 1984, ApJ, 280, 825
%-847, The Evolution and Fate of Very Massive Objects
\bibitem[\protect\citeauthoryear{Bouquet, Feix, Fualkow \& Munier}
{1985}]{bouquet1985} Bouquet S., Feix M. R., Fualkow E., Munier A.,
1985, ApJ, 293, 494
%Density Bifurcation in a homogeneous isotropic collapsing star
%
%\bibitem[\protect\citeauthoryear{CaiShu}{2005}]{caishu05}
%Cai M. J., Shu F. H., 2005, ApJ, 618, 438
%%-450; Collapse of Singular Isothermal Spheres to Black Holes.
%
\bibitem[\protect\citeauthoryear{Brown \& Bethe}{1994}]{BB94}
Brown G. E., Bethe H. A., 1994, ApJ, 423, 659
%--664; A Scenario for a Large Number
%of Low-Mass Black Holes in the Galaxy
\bibitem[\protect\citeauthoryear{Burrows}{2000}]
{adamburrows00}Burrows A., 2000, Nature, 403, 727
%--733; Adam; Supernova explosions in the Universe.
\bibitem[\protect\citeauthoryear{Chandrasekhar}{1957}]
{chandrasekhar57}Chandrasekhar S., 1957, Stellar Structure. Dover
Publications, New York
%
\bibitem[\protect\citeauthoryear{Cheng}{1978}]{cheng78}
Cheng A. F., 1978, ApJ, 221, 320
%Unsteady hydrodynamics of spherical gravitational collapse
%
\bibitem[\protect\citeauthoryear{Fatuzzo, Adams \& Myers}{2004}]{fatuzzo2004}
Fatuzzo M., Adams F. C., Myers P. C., 2004, ApJ, 615, 813
%
\bibitem[\protect\citeauthoryear{Fillmore \& Goldreich}{1984}]{fillmore84}
Fillmore J. M., Goldreich P., 1984, ApJ, 284, 1
%Self-similar gravitational collapse in an expanding universe
%
\bibitem[\protect\citeauthoryear{Foster \&
Chevalier}{1993}]{fosterchevalier93} Foster P. N., Chevalier R.
A., 1993, ApJ, 416, 303
%Gravitational Collapse of an isothermal sphere
%
\bibitem[\protect\citeauthoryear{Goldreich
\& Weber}{1980}]{goldreich80} Goldreich P., Weber S. V., ApJ,
1980, 238, 991
%Homologously collapsing stellar cores
%
%\bibitem[\protect\citeauthoryear{Hanawa}{1997}]{hanawa97}
%Hanawa T., Nakayama K., 1997, ApJ, 484, 238
%%Hanawa, Tomoyuki; Nakayama, Kunji
%%Stability of Similarity Solutions for a Gravitationally Contracting
%%Isothermal Sphere: Convergence to the Larson-Penston Solution
%%
%\bibitem[\protect\citeauthoryear{Hanawa \&
%Matsumoto}{1999}]{hanawa99} Hanawa T., Matsumoto T., 1999, ApJ,
%521, 703
%%Hanawa, Tomoyuki; Matsumoto, Tomoaki
%%Growth of a Bar Perturbation during Isothermal Collapse
%%
%\bibitem[\protect\citeauthoryear{Hanawa \&
%Matsumoto}{2000}]{hanawa00} Hanawa T., Matsumoto T., 2000, PASJ,
%52, 241
%%Hanawa, Tomoyuki; Matsumoto, Tomoaki
%%Stability of a Dynamically Collapsing Gas Sphere
\bibitem[\protect\citeauthoryear{Herrero et al.}{1992}]{herrero92}
Herrero A., Kudritzki R. P., Vilchez J. M., Kunze D., Butler K.,
Haser S., 1992, A\&A, 261, 209
%-234, Intrinsic Parameters of Galactic Luminous OB Stars
\bibitem[\protect\citeauthoryear{Hirata et al.}{1987}]{hirata87}
Hirata K. S., et al., 1987, Phys. Rev. Lett., 58, 1490
%Kamiokande II collaboration 12 neutrinos from SN 1987A.
\bibitem[\protect\citeauthoryear{Hu, Shen, Lou \& Zhang}{2006}]{HSLZ2006}
Hu J., Shen Y., Lou Y.-Q., Zhang S.N., 2006, MNRAS, 365, 345
%--351; Forming Supermassive Black Holes by Accreting Dark and Baryon Matter
%
\bibitem[\protect\citeauthoryear{Hunter}{1977}]{hunter1977}
Hunter C., 1977, ApJ, 218, 834
%
\bibitem[\protect\citeauthoryear{Hunter}{1986}]{hunter1986}
Hunter C., 1986, MNRAS, 223, 391
%\bibitem[\protect\citeauthoryear{Larson}{1969}]{larson1969}
%Larson R. B., 1969, MNRAS, 145, 271
%
\bibitem[\protect\citeauthoryear{Jordan \& Smith}{1977}]
{jordansmith1977}Jordan D. W., Smith P., 1977, Nonlinear Ordinary
Differential Equations, Oxford University Press. Oxford
%
\bibitem[\protect\citeauthoryear{Kennel \& Coroniti}{1984}]{KC1984}
Kennel C. F., Coroniti F. V., 1984, ApJ, 283, 694
%--709;
%Confinement of the Crab pulsar's wind by its supernova remnant
%
%\bibitem[\protect\citeauthoryear{KennelCoroniti}{1984b}]{KC1984b}
%Kennel C. F., Coroniti F. V., 1984b, ApJ, 283, 710
%%--730;
%%Magnetohydrodynamic model of Crab nebula radiation
%%
%%
\bibitem[\protect\citeauthoryear{Landau \& Lifshitz}{1959}]
{landau1959}Landau L. D., Lifshitz E. M., 1959, Fluid
Mechanics, Pergamon Press, New York
%
\bibitem[\protect\citeauthoryear{Larson}{1969a}]
{larson69a}Larson R. B., 1969a, MNRAS, 145, 271
%Numerical calculations of the dynamics of collapsing proto-star
%
\bibitem[\protect\citeauthoryear{Larson}{1969b}]{larson69b}
Larson R. B., 1969b, MNRAS, 145, 405
%A model for the formation of a spherical galaxy
%
\bibitem[\protect\citeauthoryear{Lattimer \&
Prakash}{2004}]{lattimer04} Lattimer J. M., Prakash M., 2004,
Science, 304, 536
%-542, The Physics of Neutron Stars
%\bibitem[\protect\citeauthoryear{Lou}{1993}]
%{lou1993}Lou Y.-Q., 1993, ApJ, 414, 656
%%--663
%%Alfvenic fluctuations in a relativistic wind with a spiral magnetic
%%field and random magnetic field structures in the Crab Nebula
%%
%\bibitem[\protect\citeauthoryear{Lou}{1994}]
%{lou1994}Lou Y.-Q., 1994, ApJ, 428, L21
%%--L24;
%%Magnetic fields in young supernova remnants
%%
\bibitem[\protect\citeauthoryear{Lou}{2005}]
{lou2005}Lou Y.-Q., 2005, ChJAA, 5, 6
%--20; Two-fluid Dynamics in Clusters of Galaxies
\bibitem[\protect\citeauthoryear{Lou \& Shen}{2004}]
{loushen2004}Lou Y.-Q., Shen Y., 2004, MNRAS, 348, 717
%Envelope expansion with core collapse -
%I. Spherical isothermal similarity solutions
%
\bibitem[\protect\citeauthoryear{Lou}{2005}]
{lou2005}Lou Y.-Q., 2005, ChJAA, 5, 6
%--20; Two-fluid Dynamics in Clusters of Galaxies
%
\bibitem[\protect\citeauthoryear{Lou \& Gao}{2006}]
{lougao2006}Lou Y.-Q., Gao Y., 2006, MNRAS, submitted
%Similarity Shocks in Polytropic Gas
%Flows around Star-Forming Regions
%
\bibitem[\protect\citeauthoryear{McLaughlin \&
Pudritz}{1997}]{mclaughlin97} McLaughlin D. E., Pudritz R. E.,
1997, ApJ, 476, 750
%McLaughlin, Dean E.; Pudritz, Ralph E.
%Gravitational Collapse and Star Formation in
%Logotropic and Nonisothermal Spheres
%
\bibitem[\protect\citeauthoryear{Murakami, Nishihara \&
Hanawa}{2004}]{murakami04} Murakami M., Nishihara K., Hanawa T.,
2004, ApJ, 607, 879
%Murakami, Masakatsu; Nishihara, Katsunobu; Hanawa, Tomoyuki
%Self-Similar Gravitational Collapse of Radiatively Cooling Spheres
%
\bibitem[\protect\citeauthoryear{Nadyozhin \&
Razinkova}{2005}]{NadyozhinRzinkova2005} Nadyozhin D. K.,
Razinkova T. L., 2005, Astronomy Letters, 31, 695
(astro-ph/0505056)
%-705, Similarity Theory of Stellar Models and
%the Structure of Very Massive Stars

\bibitem[\protect\citeauthoryear{Nauenberg \& Chapline}{1973}]{NC73}
Nauenberg M., Chapline G., 1973, ApJ, 179, 277
%Determination of properties of cold stars in General
%Relativity by a variational method.
%maximum mass of a neutron star.

\bibitem[\protect\citeauthoryear{Ori \& Piran}{1988}]{ori88} Ori
A., Piran T., 1988, MNRAS, 234, 821
%A simple stability criterion for isothermal
%spherical self-similar flow
%
\bibitem[\protect\citeauthoryear{Penston}{1969a}]{penston69a}
Penston M. V., 1969a, MNRAS, 144, 425
%Dynamics of self-gravitating gaseous spheres-
%III. Analytical results in the free-fall of isothermal cases
%
\bibitem[\protect\citeauthoryear{Penston}{1969b}]{penston69b}
Penston M. V., 1969b, MNRAS, 145, 457
%Dynamics of self-gravitating gaseous spheres-II.
%Collapses of gas spheres with cooling and the
%behaviour of polytropic gas spheres
%
\bibitem[\protect\citeauthoryear{Press et al.}{1986}]{press86}
Press W. H., Flannery B. P., Teukolsky S. A., Vetterling W.,
1986, Numerical Recipes (Cambridge: Cambridge University Press)
%
\bibitem[\protect\citeauthoryear{Rhoades \& Ruffini}{1974}]{RR74}
Rhoades C. E., Ruffini R., 1974, Phys. Rev. Lett., 32, 324
%Maximum Mass of a Neutron Star
%
\bibitem[\protect\citeauthoryear{Schaller et al.}{1992}]{schaller92}
Schaller G., Schaerer D., Meynet G., Maeder A., 1992, A\&A Suppl.
Ser., 96, 269
%-331, New Grids of Stellar Models From 0.8 to 120 M_sun at Z=0.020 and Z=0.001
\bibitem[\protect\citeauthoryear{Sch\"onberner \&
Harmanec}{1995}]{schonberner95} Sch\"onberner D., Harmanec P.,
1995, A\&A, 294, 509
%-514, On the Absolute Brightnesses and Masses of Early-Type Stars
\bibitem[\protect\citeauthoryear{Semelin, Sanchez
\& de Vega}{2001}]{semelin01} Semelin B., Sanchez N.,
de Vega H. J., 2001, Phys. Rev. D, 63, 4005
%Self-gravitating fluid dynamics, instabilities, and solitons
%
\bibitem[\protect\citeauthoryear{Shapiro \& Teukolsky}{1983}]
{shapiro83}Shapiro S. L., Teukolsky S. A., 1983, Black Holes,
White Dwarfs and Neutron Stars, John Wiley \& Sons, Inc.
\bibitem[\protect\citeauthoryear{Shen \& Lou}{2004}]
{shenlou04}Shen Y., Lou Y. Q., 2004, ApJL, 611, L117
%--L120; shocked self-similar collapse and
%flows in star formation processes
\bibitem[\protect\citeauthoryear{Shen \& Lou}{2006}]
{shenlou06}Shen Y., Lou Y. Q., 2006, MNRAS Lett, 370, L85
(astro-ph/0605505)
%--L89; Dispersal of gaseous circumstellar
%discs around high-mass stars
\bibitem[\protect\citeauthoryear{Shu}{1977}]
{shu1977}Shu F. H., 1977, ApJ, 214, 488
%
%\bibitem[\protect\citeauthoryear{Shu, Adams \& Lizano}{1987}]
%{sal87} Shu F. H., Adams F. C., Lizano S., 1987, ARA\&A, 25, 23
%%star formation in molecular clouds:observation and theory
%%
\bibitem[\protect\citeauthoryear{Shu et al.}{2002}]{shu02}
Shu F. H., Lizano S., Galli D., Cant\'o J., Laughlin G., 2002,
ApJ, 580, 969
%Self-similar Champagne Flows in H II Regions
%Shu, Frank H.; Lizano, Susana; Galli, Daniele;
% Cant\'o, Jorge; Laughlin, Gregory
%
\bibitem[\protect\citeauthoryear{Suto \& Silk}{1988}]
{sutosilk88}Suto Y., Silk J., 1988, ApJ, 326, 527
%
\bibitem[\protect\citeauthoryear{Terebey, Shu \& Cassen}{1984}]
{tsc}Terebey S., Shu F. H., Cassen P., 1984, ApJ, 286, 529
%The collapse of the cores of slowly
%rotating isothermal clouds
%
\bibitem[\protect\citeauthoryear{Tsai \& Hsu}{1995}]{th95}
Tsai J. C., Hsu J. J. L., 1995, ApJ, 448, 774
%Protostellar Collapse with a Shock
%
\bibitem[\protect\citeauthoryear{Wang \& Lou}{2006}]{WL06}
Wang W. G., Lou Y.-Q., 2006, MNRAS, submitted
%Self-Similar Dynamics of a Magnetized Polytropic Gas
%
\bibitem[\protect\citeauthoryear{Whitworth \& Summers}{1985}]
{whitworthsummers1985}Whitworth A., Summers D., 1985, MNRAS, 214, 1
%Self-similar condensation of spherically symmetric
%self-gravitating isothermal gas clouds
%
\bibitem[\protect\citeauthoryear{Yahil}{1983}]{yahil83}
Yahil A., 1983, ApJ, 265, 1047
%Self-similar stellar collapse
%
\bibitem[\protect\citeauthoryear{Yu \& Lou}{2005}]
{YuLou05}Yu C., Lou Y.-Q., 2005, MNRAS, 364, 1168
\bibitem[\protect\citeauthoryear{Yu et al.}{2006}]
{YuLou06}Yu C., Lou Y.-Q., Bian F. Y., Wu Y.,
2006, MNRAS, 370, 121 (astro-ph/0604261)
%--140; Envelope Expansion with Core Collapse. III.
%Similarity Isothermal Shocks in a Magnetofluid;
%very positive referee report; response back to MNRAS
%Editorial office a few days ago. April 11, 2006
\end{thebibliography}
\end{document}